\documentclass{ieeeaccess}
\usepackage{amsmath,amssymb,amsfonts}
\usepackage{algorithmic}
\usepackage{graphicx}
\usepackage{textcomp}
\usepackage[utf8]{inputenc}
\usepackage[T1]{fontenc}
\usepackage{microtype}
\usepackage{nccmath}
\usepackage{newtxmath}
\usepackage{mathtools}
\usepackage{hyperref}
\usepackage{float}
\usepackage{cite}
\usepackage[inline]{enumitem}
\usepackage[caption=false,font=footnotesize]{subfig}
\usepackage[export]{adjustbox}
\usepackage{physics}

\usepackage{helvet}

\newcommand{\EV}[1]{\vmathbb{E}\left[#1\right]}

\newcommand{\phm}{\phantom-}

\def\BibTeX{{\rm B\kern-.05em{\sc i\kern-.025em b}\kern-.08em
    T\kern-.1667em\lower.7ex\hbox{E}\kern-.125emX}}

\begin{document}

\history{Date of publication xxxx 00, 0000, date of current version xxxx 00, 0000.}
\doi{XXXX}

\title{A Linear Algebraic Framework for Dynamic Scheduling Over Memory-Equipped Quantum Networks}
\author{\uppercase{Paolo Fittipaldi}\authorrefmark{1},
\uppercase{Anastasios Giovanidis\authorrefmark{1,2}, and Frédéric~Grosshans}\authorrefmark{1}}
\address[1]{Sorbonne Université,  CNRS, LIP6, F-75005 Paris, France (email: \{paolo.fittipaldi, anastasios.giovanidis, frederic.grosshans\}@lip6.fr) }
\address[2]{Ericsson AI Research \& Systems, Paris, France}

\tfootnote{We acknowledge funding by the French state through the \emph{Programme d’Investissements d’Avenir} managed by the Agence Nationale de la Recherche (project ANR-21-CMAQ-0001)
and by the European Union’s \emph{Horizon 2020 research and innovation program} under
Grant Agreement No.\ 820445 and project name
\emph{``Quantum Internet Alliance''}.
This work extends \cite{FittipaldiScheduling}, which was presented at the IEEE International Conference of Quantum Computing and Engineering 2022.}

\corresp{Corresponding author: Paolo Fittipaldi (email: paolo.fittipaldi@lip6.fr).}

\begin{abstract}
Quantum Internetworking is a recent field that promises numerous interesting applications, many of which require the distribution of entanglement between arbitrary pairs of users. 
This work deals with the problem of scheduling in an arbitrary entanglement swapping quantum network 
--- often called first generation quantum network ---
in its general topology, multicommodity, loss-aware formulation.
We introduce a linear algebraic framework that exploits quantum memory through the creation of intermediate entangled links. 
The framework is then employed to apply Lyapunov Drift Minimization (a standard technique in classical network science) to mathematically derive a natural class of scheduling policies for quantum networks minimizing the square norm of the user demand backlog.
Moreover, an additional class of Max-Weight inspired policies is proposed and benchmarked, 
reducing significantly the computation cost at the price of a slight performance degradation. 
The policies are compared in terms of information availability, localization and overall network performance through an ad-hoc simulator that admits user-provided network topologies and scheduling policies in order to showcase 
the potential application of the provided tools to quantum network design. 
\end{abstract}


\titlepgskip=-15pt

\maketitle

\setlength{\marginparwidth}{\columnwidth}
\section{Introduction}
\PARstart{A}{s} experimental demonstrations of quantum repeater links and small-scale quantum networks\cite{Bernien3Metres,LagoRiveraLink,HermansNetwork} start to surface, the vision of a future Quantum Internet moves closer to reality\cite{rfc9340,WehnerQInternetVision,CacciapuotiInternet,VanMeterArchitecture}.

Despite it still being a long-term goal, the road is partially paved by the development of the classical internet, that identified and solved all the problems intrinsic to scaling a network up and operating it in a distributed way. The solutions to such problems are not directly translatable to quantum networks in general because quantum hardware is fundamentally different, creating the need for a new branch of network science with its own set of specialized tools. The present work aims to describe a novel framework to formulate and solve the problem of scheduling entanglement swapping operations in quantum networks, and showcase its potential through some application examples.

In classical networks, communication is achieved by making information packets hop through a series of network nodes until they reach their destination. Whenever several packets from different users need to pass through the same node, the node needs to have a specific discipline that regulates the order in which the packets are relayed. Depending on the application, the network might want to minimize all wait times, prioritize the packets that have certain properties or use more sophisticated specialized algorithms to determine the order of passage. The set of rules that a node applies to solve this problem is called a \emph{scheduling policy}, and it is an integral part of every well-functioning network architecture\cite{BonaldScheduling}. 

Switching to quantum networks, the concept of packet going from a source to a destination no longer applies. The cornerstone of a large and varied set of communication applications\cite{ProtocolZoo,CaleffiDQC} in the quantum domain is quantum entanglement, and the ultimate task of a quantum network system is to distribute entanglement to arbitrary sets of users.
Due to the difficulties that come with distributing entanglement over a long link, the task is achieved in practice through entanglement swapping operations at intermediate nodes
\cite{AzumaRepeaters}
that may serve several distinguished pairs of end users.
The challenge of scheduling in quantum networks revolves therefore around entanglement swapping operations, which must be scheduled by the nodes following what will be addressed in the following as a \emph{quantum scheduling policy}.

Despite there being several solutions that yield an extensive choice of well-established policies for classical networks, the scheduling problem remains an active challenge for quantum networks: pioneeristic effort has been undertaken to solve the scheduling problem in specific quantum networking settings\cite{SkrzypczykArchitecture,DaiScheduling,NguyenMultiRouting,ChandraScheduling,DaiTowsleyScheduling}, but no trivial generalization of the results presented in these works to medium and large scale networks is possible. 

In this context, our work aims to provide a framework that can be employed for designing and benchmarking scheduling policies on general quantum networks. We stress that our findings pertain to arbitrary network topologies with no theoretical limit on scale and enable users to work with multiple commodities requesting streams of entangled pairs. Furthermore, our framework actively exploits quantum memory slots: even when not all elementary links along a given route are ready, the network is still allowed to create intermediate entangled pairs that cover a part of the route exploiting the available links and store them in memory for future use. The idea of intermediate links has already appeared in other works\cite{RoutingMemoriesWehnerKerenidis,ChakrabortyDistributedRouting,Pouryousef2022QON}, and we seek to extend it to our general setting as a core mechanism of operation of the network systems we model. 

It should be noted that, while some scheduling policies are proposed and analysed in the following, the broader focus of this work is on describing the framework as a practical tool and providing examples of its application to non-trivial scenarios.

Our work is primarily aimed at first generation quantum networks as detailed in \cite{1g2g3g}, but our methods might prove interesting for a future treatment of second and third generation systems as well.

The paper follows the following structure: in sec. \ref{sec_context}, the relevant scientific literature is reviewed and compared with our contribution. Sec. \ref{sec_sysdesc} provides a detailed description of the system we are modeling and the various components of our algebraic framework. We follow up with sec. \ref{sec_policies}, where we introduce and analyze an array of scheduling policies through the tools we propose. Sec. \ref{sec_numerical} is devoted to presenting numerical results obtained by applying our tools to several network setups. 
\section{Context and Relevance of This Work}
\label{sec_context}
As a cross-disciplinary topic, quantum networks are interesting to both quantum physicists and classical network scientists: many protocol-level questions already have an established answer in the classical domain, but the fundamental differences specific to quantum hardware require careful reconsideration of the pre-existing knowledge base. Much like our work, \cite{SkrzypczykArchitecture} provides a formulation of the scheduling problem on quantum networks, the main difference being that the cited work approaches the problem through architecture design and heuristic scheduling, whereas our contribution is more geared towards building a general algebraic framework to mathematically derive and compare scheduling policies. 

Concerning purely theoretical results, an optimal theoretical bound for entanglement distribution across a line network with a single commodity is derived in \cite{DaiScheduling} and expanded upon in \cite{DaiTowsleyScheduling}.

References \cite{VardoyanSwitchStochastic},\cite{DaiSwitchProtocols}, and \cite{ChandraScheduling} are all examples of stochastic analysis of a single quantum switch to characterize the scheduling policies that stabilize it. The physical model employed in these works is deeper, in that it accounts for purely quantum imperfections that we neglect, but their scope is somewhat narrower than ours because they all consider a single quantum switch that has to serve a set of users in a star-like configuration. 

More specifically relevant to our work, \cite{VasantamSwitch} and \cite{PanigrahySwitch} detail the application of Lyapunov stability theory to a quantum switch and the subsequent derivation of a throughput-optimal Max Weight \cite{Tassiulas} policy, much like it is done for the quadratic policies we propose. The key differences rely in the generality of our work, which applies to arbitrary topologies with multiple commodities, and in the fact that the cited papers model a switch as a single-hop queuing system dealing with entanglement requests, i.e.\@ requests arrive at the switch and are served after waiting in a queue. 
Here, we extend this idea by also keeping track of entangled qubit pairs (referred to as an \emph{ebit}\cite{wilde} hereafter) through a multi-hop  queueing model: 
generation and usage of ebits are naturally modeled through simple enqueuing and dequeuing, while entanglement swapping operations entail simultaneous dequeuing from two parent queues and enqueuing in a child one. 

This new set of queues for ebits acts as a variable resource that the network must regulate according to a suitable scheduling policy. The two queuing models are described in more details in sec.\@ \ref{sec_sysdesc}.

The usage of memory in our framework is physically similar to the Virtual Quantum Link idea first introduced in \cite{RoutingMemoriesWehnerKerenidis} and revisited in \cite{ChakrabortyDistributedRouting, DaiTowsleyScheduling, Pouryousef2022QON}: the introduction of memory at the nodes enables them to seek a balance between depleting their supply of entangled pairs for swapping and conserving it for future use or direct user consumption. The deeper implication of this point is that the network is free to create intermediate links and store them: this leads to distributing pairs across a service route in a ``growing'' fashion, that both increases performance and removes the need for end-to-end link state information, while naturally adapting to a multi-hop queuing scenario.

As a final remark, we stress that due to the abundance of interesting research that has been carried out to perform quantum routing on several network topologies\cite{ChakrabortyDistributedRouting,CoutinhoRouting,GyongyosiRouting,NguyenMultiRouting} we assume the existence of a set of static pre-computed routes that connect each end-users pair, under the premise that our work should be easily integrable with a more refined routing technique.

To conclude the section, we summarize the key contributions of the present manuscript:
\begin{enumerate}
    \item We introduce a general framework for scheduling in quantum networks that poses no assumptions on topology, number of commodities or choice of scheduling policy (sec. \ref{sec_sysdesc});
    \item We extend the idea of intermediate virtual link to the general network case (\textit{ibidem});
    \item Through the help of our framework, we derive a throughput-efficient quadratic scheduling policy that works over our multi-hop model. We then formulate different versions of this policy that relax information requirements (sec. \ref{sec_sub_quadsched});
    \item Finally, we propose a novel, Max-Weight inspired class of scheduling policies that is shown to perform satisfactorily while posing feasible communication constraints on the network (\textit{ibidem}). 
\end{enumerate}

\section{System Description}
\label{sec_sysdesc}
In this section, we describe the physical model that we will rely on to develop our framework. Since the framework we provide is composed of two interconnected queuing models, we devote subsections \ref{sec_sub_EbitQueues} and \ref{sec_sub_DemandQueues} to describe respectively the details behind \emph{ebit queues} and \emph{demand queues}.
As a preliminary step, we clarify the notation conventions that are adopted in this work: lower case for scalars ($x$), bold lower case for vectors ($\mathbf{x}$), bold upper case for matrices ($\mathbf{X}$) and calligraphic upper case for sets ($\mathcal{X}$). Well-known matrices such as the identity matrix or the null matrix are indicated in blackboard bold and subscripted with their dimension, as in $\vmathbb{I}_{n}$ and $\vmathbb{0}_{n\times m}$.

Since the term is ubiquitous in the following, we state the definition of a quantum switch
as a device that is equipped with quantum memories to store qubits, a Bell State Measurement (BSM) apparatus to perform entanglement swapping operations, and local quantum processing capabilities. An entanglement swapping operation is assumed to be instantaneous and always successful, and the classical communication overhead that comes with entanglement swapping (such as sharing measurement results) is considered free. 
We assume our quantum switches to be connected to a classical communication infrastructure to coordinate control operations for protocols and, if the chosen scheduling policy so requires, exchange status information with other nodes and/or a central scheduling controller. Moreover, every node is assumed to possess unlimited memory slots. While this might look like too coarse of an assumption, both the literature\cite{VardoyanSwitchStochastic,NainSwitch} and some preliminary results we present here suggest that, while indeed being an important modeling point, limiting the memory slots might not be the first network limitation that must be taken into account.  

The physical system we consider is a network of quantum switches connected by lossy fiber links. 
We model it as an arbitrary connected graph $\mathcal{G} = (\mathcal{V},\mathcal{E})$, where the switches are deployed at the locations specified by the vertices $\mathcal{V}$ and interconnected by edges $(i,j) \in \mathcal{E}$ that represent a fiber link plus a generic elementary entangement generation scheme (such as a $\chi^{(2)}$ crystal or a Bell State Analyzer in the middle\cite{DLCZ} or at one of the stations\cite[sec. V-C]{AzumaRepeaters}).
Every switch has a number of memory slots, assumed to be infinite in this work, in which qubits may be stored. Ebits are generated by each fiber link with a given constant average rate, which may be heterogeneous across links but is constant in time, and stored inside memories at the end nodes of the respective link. Among the network nodes, 
there are $n$ pairs
$\{(\mathit{Alice}_1, \mathit{Bob}_1),\ldots,(\mathit{Alice}_n, \mathit{Bob}_n)\}$
that request ebits in a random way to realize a generic application. Each $(\mathit{Alice}_n, \mathit{Bob}_n)$ pairs is connected by one or more routes that are not necessarily disjoint form the ones connecting other users, and therefore can create congestion that needs to be managed by a scheduling policy. We stress that since we assume unlimited memory we are choosing to focus on the link congestion case: we leave node congestion for future investigation.

Given this starting point, the purpose of a quantum network is to perform entanglement swapping operations in order to distribute ebits to its users in a way that maximizes a given performance metric. In pursuing this objective, the network must rely on a scheduling policy to minimize congestion by carefully deciding which swaps to perform when, while also being hindered by link-level fiber losses and by quantum memory imperfection causing the loss of stored ebits. 

Memory and fiber losses are the only two sources of imperfection that are accounted for in this paper: for simplicity reasons, we neglect  sources of state degradation other than losses in this formulation of our algebraic model, since they require a far lower level of abstraction, and lead to more complex multiobjective problems \cite{bugalho,ChakrabortyMulticommodity}. However, our model could be reinterpreted in the context of more modern error-corrected networks if we state that each link generates entangled pairs with a given \emph{logical rate}, i.e. the rate of creation of error-corrected ebits.

For practical reasons, we discretize the time axis: since the scheduler is supposed to take decisions at fixed times, it is natural to take a discrete time step $\Delta t$ as the time unit of interest. 

Between two subsequent clock ticks the system is free to evolve stochastically and at the end of each time step a scheduling decision is taken. This places a lower bound on $\Delta t$: no decision can happen before all information has been successfully communicated to all deciding agents, thus $\Delta t$ must be at least as large as the classical communication delay introduced by state-related information exchange. We note that, while at the moment our work does not take into account finite communication delays, the design process of a real system would need to consider that a policy that requires more communication, despite being better informed, will suffer from more losses (as they depend on the length of the time step) and be less reactive to instantaneous change. 

\subsection{Ebit Queues}
\label{sec_sub_EbitQueues}
To model ebits stored at memory nodes, the concept of an \emph{ebit queue} is introduced: each pair of nodes $e = (i,j)$ inside the extended edge set $\tilde{\mathcal{E}}=\mathcal{V}\times\mathcal{V}$ is said to possess an ebit queue $q_{ij}(t)$. Furthermore, among ebit queues, every $q_{ij}(t)$ associated to an edge $(i,j)\in\mathcal{E}$ corresponds to an elementary entanglement generation link, and is therefore called a \emph{physical queue}, whereas all other ebit queues are called \emph{virtual queues}. Ebit queues are therefore a piece of classical control information introduced to keep track of which nodes share entanglement: $q_{ij}(t) = n$ means that there are $n$ qubits at node $i$ and $n$ qubits at node $j$, taking up $n$ memory slots at the respective nodes and sharing pairwise entanglement. 
In the following, we describe how all the processes that ebits undergo in our model are translated to queue operations.
\subsubsection{Ebit Generation}
At each time step, every fiber link --- and thus every physical queue --- generates a random number of 
ebits $a_{ij}(t)$. This term can be seen as an open interface to the specific random process that models ebit generation and it is treated hereafter as a Poisson process --- which in our discrete model becomes a Poissonian random variable ---. 
For the sake of simplicity, we assume $a_{ij}(t)$ to have constant mean value $\alpha_{ij}\geq0$, i.e.\@ at every time step, $\alpha_{ij}$ ebits are generated on average. It should be noted that $\alpha_{ij}$ is the final generation rate after accounting for link-level imperfections --- finite brightness of the source, propagation losses, finite
success probability of pair-generation BSMs, etc.\@ --- as a cascade of Poisson filtration processes, at the end of which we obtain a value for $\alpha_{ij}$.
Thus, ebit generation is modeled by a direct enqueueing operation along the relevant queue. It should be noted that, since this operation models entanglement generation at the physical level, it only concerns physical queues. For virtual queues, $a_{ij}(t) = 0\,\forall\,t$.
\subsubsection{Ebit Losses}
To model (symmetrical) memory loss, we employ a standard quantum memory model and calculate the storage-and-retrieval efficiency of the memories as $\eta = \exp{\left(-\frac{\Delta t}{\tau}\right)}$, where $\tau$ is the expected lifetime of a qubit in the memory and $\Delta t$ is the duration of a time step. This figure of merit models the probability to correctly retrieve a qubit from a memory after it has been stored in it for one time step. We assume losses to be symmetrical in that whenever one loss event happens, either both ends of the link lose their respective qubit or one end loses it and instantly communicates loss to the other concerned node. Therefore, one loss event always models the loss of one complete ebit.

At every time step, every queue throws as many biased coins as there are stored qubits and removes as losses all the ones that fail the random check. Losses are therefore modeled by the binomially distributed random variable $\ell_{ij}(t)$, with as many trials as there are ebits stored in queue $(i,j)$ and probability to lose one pair $1 - \eta$. It should be clear that the number of trials for the geometric distribution is based on $q_{ij}(t)$, i.e. on the pairs present at the beginning of the time step, meaning that new arrivals are immune to losses for the ongoing time step. 

We remark that the statistical distribution of ebit survival times follows the geometric distribution defined by $\eta$, whose mean value $\tfrac{1}{1-\eta}$ tends to the expected $\tfrac{\tau}{\Delta t}$ for small $\frac{\Delta t}{\tau}$, $\tau$ being the expected lifetime of ebits in the memories. The remaining difference is an effect of the dicretization. Finally, we stress that accounting for losses in such a time-dependent way makes the presented framework valid as a tool to determine the optimal frequency at which scheduling decision should be taken, given the technological parameters. 

\subsubsection{Entanglement Swapping}
After covering generation and loss, the last mechanism that can modify the amount of ebits in a queue is entanglement swapping. Entanglement swapping always involves consuming two "shorter" pairs to obtain one longer pair, which naturally translates to our queue-based formalism as two removals from the parent queues, and one addition to the child queue. We introduce the following notation: let $r_{i[j]k}(t)$ indicate the number of swapping operations that happen at a given time step, at node $j$, from queues $(i,j)$ and $(j,k)$ to queue $(i,k)$: as a notation example, $r_{A[B]C}(2) = 3$ means that the scheduler has ordered three BSMs to be performed at node $B$ to swap three pairs from queues $AB$, $BC$ to $AC$ at time step $2$. There will be as many $r_{i[j]k}(t)$ terms as there are transitions allowed by the chosen routing: if for instance there are two parallel paths $ABCD$ and $AB'C'D$ across the Alice-Bob pair $AD$, but only $ABCD$ is explicitly routed, the system will include terms $r_{A[B]C}(t)$ and $r_{A[C]D}(t)$, but not $r_{A[B']C'}(t)$ and $r_{A[C']D}(t)$, effectively ignoring the second path. This is a limitation that directly arises from assuming that routing is static and known, but is also easily circumvented by adding more paths to the routing, since we place no theoretical limit on the number of routes that can serve a user pair.

To clarify how all the pieces introduced until now fit together, suppose to have the Alice-Bob pair $AD$ connected by route $ABCD$, as shown in Fig.~\ref{fig_simexample}. Assume the average generation rates to be $\alpha_{AB}$, $\alpha_{BC}$ and $\alpha_{CD} = 1\  \text{(time steps)}^{-1}$. Lastly, assume that all the memories in the system have $\eta = 0.9$ storage-and-retrieval efficiency for the chosen time step duration. Fig. \ref{fig_simexample} shows how the full system evolves throughout two time steps, while Fig. \ref{fig_simtiming} shows the same test run but focusing on queue $AB$, to highlight the timing of the various phenomena at play.

\Figure[!t]()[width=.9\linewidth]{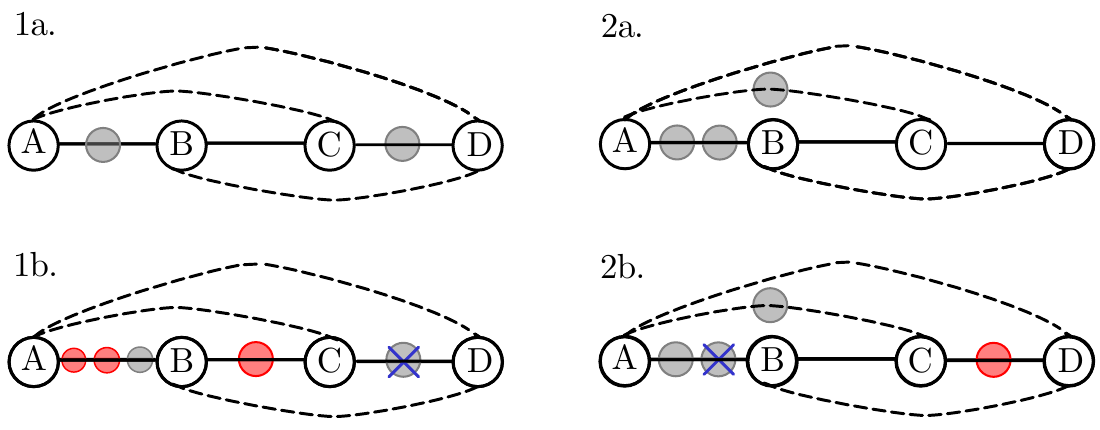}
   {Explicit example of two successive time steps over a simple chain topology. Continuous lines represent physical links, and therefore correspond to physical queues in our model, whereas dashed lines symbolize virtual queues, i.e. pairs of nodes that are not directly connected by fiber but may share ebits after some entanglement swapping operations. 
   Grey circles represent ebits in the queue at the beginning of the current time step. 
   Their number is encoded in $\mathbf{q}(t)$ in our model.
   Red circles represents ebits arrived during that time step ($\mathbf{a}(t)$). 
   Blue crosses represent loss of an ebit ($\boldsymbol{\ell}(t)$). 
   Upper figures (a) represent the state at the beginning of the corresponding time step, lower figures (b) at the end of it. Passing from time step $1$ to $2$, the scheduling decision $r_{A[B]C}(t=1) = 1$ has been applied which removed one ebit each from queues $AB$ and $BC$ and added one to $AC$.\label{fig_simexample}}

\Figure[!t]()[width=.9\linewidth]{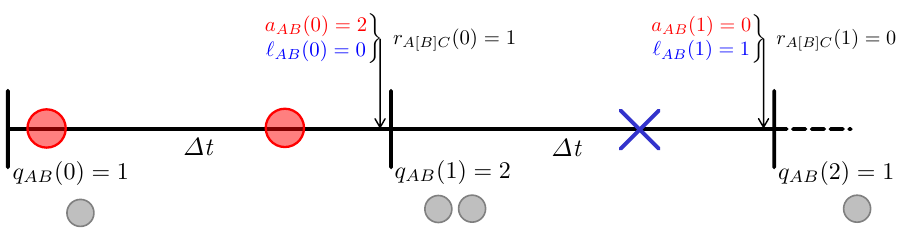}
   {Same example as fig. \ref{fig_simexample}, as seen internally by queue $AB$ to highlight the timing of the various phenomena at play. Queue snapshots $q_{ij}(t)$ are taken at the very beginning of a time step, whereas arrivals and losses happen stochastically. At the end of each time step, arrivals and losses are counted and the scheduling decision is taken. Note that ebits arriving during the current time step are not subject to losses in this model.\label{fig_simtiming}}

$\bullet$ During time step 1:
\begin{enumerate}
    \item At the beginning of the time step (fig.\ref{fig_simexample}, $1a$), the queue states are: $q_{AB}(1) = q_{CD}(1) = 1$, $q_{BC}(1) = 0$;
    \item At the end of the time step (fig.\ref{fig_simexample}, $1b$), new ebits have been generated across $AB$ and $BC$ ($a_{AB}(1) = 2$, $a_{BC}(1) = 1$, red circles) and one has been lost across $CD$ ($\ell_{CD}(1) = 1$, crossed-out grey circle). The scheduling decision is taken from this configuration as $r_{A[B]C}(t=1) = 1$: one swap at node $B$ with queues $AB$ and $BC$ as parents and $AC$ as child. The time step is concluded by the application of the scheduling decision.
\end{enumerate}

$\bullet$ During time step 2:
\begin{enumerate}
    \item The initial configuration (fig.\ref{fig_simexample}, $2a$) sees two stored pairs in $AB$ which were not employed in the last time step ($q_{AB}(2) = 2$) and the freshly swapped one in $AC$ ($q_{AC}(2) = 1$);
    \item At the end of the time step (fig.\ref{fig_simexample}, $2b$), one pair was lost across $AB$ ($\ell_{AB}(2) = 1$) and one generated across $CD$. The scheduler may now decide $r_{A[C]D}(2) = 1$ to move to $AD$ or store the pairs for future use.
\end{enumerate}

To categorize transitions in terms of their net effect on queues, we say that a given transition $i[j]k$ is
\textit{incoming} for queue $(i,k)$, because it adds pairs to it, and \textit{outgoing} for queues $(i,j)$ and $(j,k)$, because it takes pairs from them. A queue's evolution can therefore be summarized as follows, i.t.\@ and o.t.\@ being shortcuts for incoming and outgoing transitions:
\begin{multline}
    q_{ij}(t+1) = q_{ij}(t) + a_{ij}(t) - \ell_{ij}(t) \\ - \smashoperator{\sum_{o \in \text{o.t.}}}r_o(t) + \smashoperator{\sum_{k \in \text{i.t.}}}r_k(t). 
    \label{scalarevolQ}
\end{multline}
For clarity, we reiterate that while all terms of (\ref{scalarevolQ}) are calculated for every queue, $a_{ij}(t)$ across a virtual queue will always be zero, because virtual queues do not generate ebits. Moreover, it is quite rare for a physical pair to have incoming transitions, but not impossible: it may happen in a peculiar topology such as the $ABC$ triangle with $AB$ as an Alice-Bob pair and $ACB$ as service route. In this edge case, transition $A[C]B$ is incoming for a physical queue. 

Conversely, it should be stressed that the loss term $\ell_{ij}(t)$ is calculated in the same way for all queues, because ebit storage is always handled by memories at the network nodes.

\subsubsection{Vector Formulation}
A description of the whole system requires $|\tilde{\mathcal{E}}|$ equations like (\ref{scalarevolQ}), ushering a natural transition to a model built with matrices and vectors. 

The first vector terms are $\mathbf{q}(t)$, $\mathbf{a}(t)$ and $\boldsymbol{\ell}(t)$, whose $N_{\text{queues}}$ entries correspond to the individual $q_{ij}(t)$, $a_{ij}(t)$ and $\ell_{ij}(t)$ values (the ordering is irrelevant as long as it is consistent). Moreover, since the effect of swapping on the queues is linear, it is possible to describe it by introducing the vector $\mathbf{r}(t)$, which has $N_{\text{transitions}}$ elements --- and a matrix $\mathbf{M}$ with $N_\text{queues}$ rows and $N_\text{transitions}$ columns to translate the transition rates into their net effect on queues. 

The $\mathbf{r}(t)$ vector embodies the scheduling decision and it is a mere list of all the $r_{i[j]k}$ terms, while the $\mathbf{M}$ matrix introduces an efficient encoding of the network topology and routes: For each of its columns, associated to transition $i[j]k$, the $\mathbf{M}$ matrix has $-1$ on the rows associated to queues $(i,j)$ and $(j,k)$, and $+1$ on the row associated to queue $(i,k)$. All other terms are zero. An example of the $\mathbf{M}$ matrix is given in table \ref{tab:Mexample} in order to provide the reader with intuition on how it's built. We remark that in all non-trivial examples that are analyzed in this work the $\mathbf{M}$ matrix is automatically generated by our simulator.

System-wide queue evolution can be restated as the following simple linear equation:
\begin{align}
\mathbf{q}(t+1)  & = \mathbf{q}(t) - \boldsymbol\ell(t) + \mathbf{a}(t) + \mathbf{Mr}(t).
\label{vectorialevolQ}
\end{align}

\begin{table}[tb]
    \centering
    \caption{$\mathbf{M}$ matrix for the linear $ABCD$ network}
    \begin{tabular}{l | c c c c}
    & ${A[B]C}$ & ${B[C]D}$ & ${A[B]D}$ & ${A[C]D}$\\\hline
    $AB$ & $-1$ & $\phm0$ & $-1$ & $\phm0$ \\
    $BC$ & $-1$ & $-1$ & $\phm0$ & $\phm0$ \\ 
    $CD$ & $\phm0$ & $-1$ & $\phm0$ & $-1$ \\
    $AC$ & $+1$ & $\phm0$ & $\phm0$ & $-1$ \\
    $BD$ & $\phm0$ & $+1$ & $-1$ & $\phm0$ \\
    $AD$ & $\phm0$ & $\phm0$ & $+1$ & $+1$ \\
    \end{tabular}
    \label{tab:Mexample}
    \vspace{-0.5cm}
\end{table}
Looking at table \ref{tab:Mexample}, notice that, as this work only involves bipartite entanglement, all columns of $M$ have two $-1$ terms and one $1$. It would be possible to generalize this model to n-party entanglement by introducing multipartite queues and defining transitions that add to them by drawing from three or more bipartite queues to model a protocol similar to the ones shown in \cite{MeignantMultipartite,PirkerMultiParty}. For the sake of simplicity and avoiding the severe scaling problems this generalization would create, we focus on bipartite states for now. This entails that every column of $M$ sums to -1, i.e. every swap operation has the net effect of removing one pair from the system.

\subsubsection{Ebit Consumption}
Up to now, the scheduler can freely swap pairs in the network but there is no mechanism for users to employ the received pairs. The missing piece of the puzzle for ebit queues is consumption: whenever there is availability of entangled pairs across one of the final $(\mathit{Alice}_n,\mathit{Bob}_n)$ pairs, the scheduler must be able to use the available pairs to serve requests, i.e. consume the distributed resource. This is implemented in the model by extending the matrix $\mathbf{M}$ through concatenation of a negative identity block to obtain $\tilde{\mathbf{M}} = \left[ \mathbf{M} \middle|-\vmathbb{I}_{N_{\text{queues}}}  \right]$, and the $\mathbf{r}(t)$  vector to have $N_{\text{transitions}} + N_{\text{queues}}$ components. 

What this extension achieves is to have a set of new transitions that only remove one pair from a given queue, modeling actual consumption of the distributed pair by the users. Extending $\mathbf{M}$ to $\tilde{\mathbf{M}}$ empowers the scheduler but also adds a new facet to the decision problem: if a given queue has $n$ pairs inside, the scheduler not only needs to balance swapping and storage for future use, it might also have to account for direct consumption of some of the available ebits. 

Putting all the terms together, the vector of ebit queues evolves as:
\begin{align}
\mathbf{q}(t+1)  & = \mathbf{q}(t) - \boldsymbol\ell(t) + \mathbf{a}(t) + \mathbf{\tilde{M}r}(t).
\label{vectorialevolQwithMtilde}
\end{align}

\subsection{Demand Queues}
\label{sec_sub_DemandQueues}
The ultimate purpose of a communication network is to serve the requests that users issue. Therefore, we need to include in our discussion a mechanism that allows to keep track of user demand: at any given time, every $(\mathit{Alice}_n, \mathit{Bob}_n)$ pair will issue a random number of demands and store them in a backlog called the \emph{demand queue}. Every time a direct consumption operation is scheduled and a pair is consumed along link $ij$, a demand is contextually removed from the demand queue of link $ij$. This physically corresponds to the users measuring their qubits and ``consuming'' one ebit to realize the specific application they are implementing.

Thus, it becomes natural to introduce another set of queues to describe the evolution of demands.  
Similarly to ebits, demands arriving to the system and being held for future service are modeled through queues: alongside every ebit queue, there exists a demand queue $d_{ij}(t)$ that keeps track of the number of user-issued requests (as introduced in \cite{DaiSwitchProtocols} for a single switch and generalized in this work for an arbitrary topology). At each time step, every demand queue $d_{ij}(t)$ receives $b_{ij}(t)$ demands, which for simplicity and generality are again modeled as a Poisson process with constant average value $\beta_{ij}$ (as in the case of ebit generation, this term may be interpreted as an open interface to more refined traffic patterns). To maintain the model's uniformity, all edges belonging to $\tilde{\mathcal{E}}$ have a demand queue, but only the ones that are associated to an $(\mathit{Alice}_n, \mathit{Bob}_n)$ pair have nonzero arrivals. For all the other links, $b_{ij}(t) = 0\,\forall\,t$.

Demand queues have a simpler evolution than ebit queues, since a demand is only a request for one ebit to be distributed across a given $(\mathit{Alice}, \mathit{Bob})$ pair: demands enter their queues when they are received and exit when they are served. Demand service can be naturally controlled by the $ij$ terms of the $\mathbf{r}(t)$ vector, i.e.\@ the same terms that control ebit consumption. We therefore introduce the matrix $\tilde{\mathbf{N}} =  \left[\vmathbb{0}_{N_{\text{queues}}\times N_{\text{transitions}}} \middle| -\vmathbb{I}_{N_{\text{queues}}}\right]$ as a mean of interfacing with the consumption part of the $\mathbf{r}(t)$ vector without being affected by the scheduling one, which is irrelevant to demand queues.

Demand evolution may therefore be stated as:
\begin{align}
\mathbf{d}(t+1) &= \mathbf{d}(t) + \mathbf{b}(t) + \mathbf{\tilde{N}r}(t)
\label{vectorialevolD}
\end{align}
By construction, the last $N_\text{queues}$ components of the $\mathbf{r}(t)$ vector regulate both demand and ebit consumption: one demand always consumes one ebit.

\section{Scheduling Policies}
\label{sec_policies}
\subsection{General Overview}
\label{sec:sched_intro}
After introducing all the components of the model, we move to describing scheduling policies and how they can be tested through our tools. 
We first outline what a scheduling policy is in the context of our work and follow up with subsections dedicated to three categories of scheduling policies: subsection \ref{sec_sub_Greedy} describes the Greedy scheduler, i.e.\@ the simplest policy we analyze in this work; subsection \ref{sec_sub_quadsched} features a mathematical derivation of a quadratic family of scheduling policies; subsection \ref{sec_sub_mwsched} shows how the quadratic schedulers can be modified to obtain a class of policies that perform similarly but require lighter computations.
We define a \textit{Scheduling Policy} as any arbitrary set of rules that at every time step $t$ takes as its input some degree of information about the network state and returns a scheduling decision $\mathbf{r}(t)$, i.e.\@ a scheduling vector as defined in the previous section. 

We first subdivide policies according to their localization degree: in \emph{distributed} policies, the nodes themselves determine the operations to perform; in centralized ones, the system features a physical scheduler to which all the nodes communicate their status information and receive orders from. It is moreover possible to categorize policies in terms of information availability: we remark that in all policies that we analyze in the following we work on the assumption that $(\mathbf{q}(t),\mathbf{d}(t))$, i.e.\@ the exact state of the system at the beginning of time step $t$, is known to all parties. However, since networks are distributed systems, it may happen that some crucial information (such as the realizations of the random processes $a_{ij}(t)$ and $\ell_{ij}(t)$ for faraway queues) is not available or outdated when the scheduling decision is taken, introducing the notion of \emph{feasibility} of a scheduling decision, which is detailed in the following section.

\subsection{Managing Infeasibility}
\label{sec:conflicts}
To start discussing how infeasible decisions are handled, let us assume a centralized scheduler, with complete access to information. 
Let $\mathcal{I}(t)$ be the set of information accessible to the scheduler at time $t$. 
For example, a fully informed scheduler will have access to the information set $\mathcal{I}^\text{FI}(t) = \left\{\mathbf{q}(t),\mathbf{d}(t),\mathbf{a}(t),\boldsymbol\ell(t),\mathbf{b}(t)\right\}$, i.e. the state $(\mathbf{q}(t),\mathbf{d}(t))$ of the system at time $t$ plus the realizations of all the random quantities at play, making it so that the scheduler perfectly knows the state of the system at the end of the time step.
Other, more realistic schedulers, will only have access to a subset of this information, as we will see later.

As shown in sec.\@ \ref{sec_sysdesc}, the net effect of a scheduling decision $\mathbf{r}(t)$ on the ebit and demand queues is given respectively by $\mathbf{\tilde{M}r}(t)$ and $\mathbf{\tilde{N}r}(t)$. We can set two bounds on the decision:
\begin{enumerate}
    \item The net number of outgoing ebits from any given queue can never exceed what is physically available:
    \begin{align}
        -\mathbf{\tilde{M}r}(t)\leq 
\mathbf{q}(t) - \boldsymbol{\ell}(t) + \mathbf{a}(t).
\label{constr_feasibilityEBIT}
    \end{align}
    \item Along a queue, the number of consumed ebits should never be higher than the demands:
    \begin{align}
    -\mathbf{\tilde{N}r}(t)\leq 
\mathbf{d}(t) + \mathbf{b}(t)
\label{constr_feasibilityDEMAND}
    \end{align}    
\end{enumerate}
We refer to those bounds as the \emph{feasibility} bounds. 

If we now suppose (as will be the case for most of the scheduling policies presented hereafter) to have incomplete access to information, one or more of the random processes' realizations become inaccessible, making it impossible to exactly formulate the feasibility bounds. Despite it still being possible to design scheduling policies that perform well while only using educated guesses based on averages, it is not possible to guarantee that their decisions at each time instant will respect (\ref{constr_feasibilityEBIT}) and (\ref{constr_feasibilityDEMAND}). 

Infeasibilities in general arise when $n$ ebits are available in a queue and $n'>n$ are scheduled out of it; they may be caused by a central scheduler relying on outdated information and scheduling more pairs than available, or by conflicts between two node-local schedulers (sec. \ref{sec:locquad}) that try to draw from the same queue. 

Infeasibile decisions themselves do not prevent a network from operating (performing more measurements than there are available ebits simply results in failure of the excess measurements), but infeasibility that is not properly managed may entail sensible degradation of performance. Therefore, a working quantum network stack also needs a specific discipline to manage infeasible orders. 
In the context of this work, conflicting requests are managed in a random order, to mimic a real network adopting a first-come-first-serve (FCFS) discipline. 

As an example, suppose to have one ebit in queue $BC$. It may happen that a scheduling policy requests $r_{A[B]C} = 1$ and $r_{B[C]D} = 1$, two operations that feature $BC$ as a parent and therefore compete for the single available ebit. At this point, one needs to choose the discipline according to which priority is assigned: whereas this basic example may be solved by simple random selection, we illustrate in the following a more in-depth example to show the full complexity of this problem and the solution we adopt in our work. 

It could happen that the scheduler ordered to feed $q_{AC}$ through $r_{A[B]C} = 1$, exploit the new $AC$ pair in $r_{A[C]D} = 1$ and finally serve one request with $r_{AD} = 1$. Each of these operations depends on the one before it, and if the execution sequence is not respected the system will serve one less $AD$ request, possibly also wasting the intermediate links in the process and ultimately degrading performance. 

Therefore, to ensure proper priority is respected, we introduce a ranking system for swapping and consumption operations to preserve execution order. Swapping transitions and consumption orders are grouped by ranks, and the ranks are executed sequentially. All conflicts inside a rank are managed through random selection. 

\begin{figure}
\includegraphics[width=.9\linewidth]{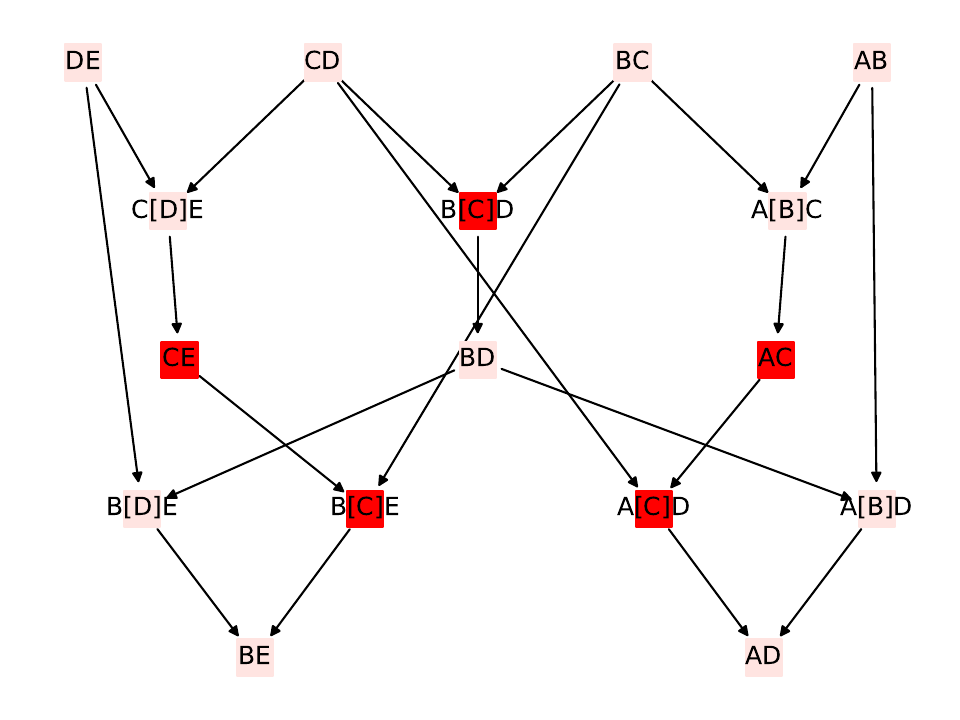}
\caption[Scheme of the rank-based conflict management system.]{The scheme of our rank system for an $ABCDE$ chain topology. Every square with only two letters inside (e.g.\@ $DE$) represents a consumption operation along a given link, while three-letter squares (e.g.\@ $C[D]E$) represent swapping transitions. A set of squares at the same height are grouped in one rank, starting from zero at the top (direct consumption from physical queues) and increasing going down. Arrows represent the ''paths`` to follow to obtain one of the final, user-requested pairs.
Focusing on the bright red squares in this scheme, which all involve node $C$ in some way, we can provide an example of how the conflict-management system works. Whenever it needs to apply a scheduling order, node $C$ will sequentially: 
\begin{enumerate*}
\item Perform transition $B[C]D$ (rank $1$) as many times as requested;
\item Satisfy consumption orders along $CE$ and $AC$ in a random order, since they are competing rank $2$ operations;
\item Perform transitions $B[C]E$ and $A[C]D$ in a random order, since they are competing rank $3$ operations.
\end{enumerate*}

As discussed in the main text, $CE$ and $AC$'s consumption orders are satisfied in a random order, but always after the upstream $B[C]D$ transition and always before the downstream $B[C]E$ and $A[C]D$ transistions.
\label{fig:RankSystem}}
\end{figure}

To form the ranks, we start by assigning rank $0$ to consumption orders from physical queues: these orders will be executed first. Secondly, swapping operations whose parents are physical queues are assigned rank $1$, so that they are executed second. After that, ranks are assigned in iterative order: for $n=0,1,2\ldots$, even ranks $2n$ are assigned to consumption orders along queues that are fed by transitions of rank at most $2n - 1$, and odd ranks $2n+1$ to transitions whose parent queues have rank at most $2n$, as depicted in fig. \ref{fig:RankSystem}. Such a system allows for a good balance between fair management of conflicting requests while preserving the sequentiality of ``ascending'' entanglement distribution operations and is also easily implemented in practice: we assume that every node knows the rank of the queues and transitions that involve it, and we envision to send, instead of a single ``apply decision'' control signal, a series of ``apply rank $n$ operations'' signals. To help intuition, an example of how the ranking system works is provided and commented in fig. \ref{fig:RankSystem}.

In the following sections we propose some examples of scheduling policies and provide details on their degree of localization and information availability. 

\subsection{Greedy Scheduler}
\label{sec_sub_Greedy}
The Greedy Scheduler is a nontrivial, distributed scheduling policy that works with minimal communication between the nodes. It is a natural and immediate solution to the scheduling problem, and it is commonly found in classical network literature as a test case. Under a greedy scheduling policy, all nodes perform swapping operations as soon as they are available, regardless of user demand. When several competing operations are available, the node selects randomly. It should be noted that, although it disregards user demand, the greedy scheduler we examine is still \emph{routing-aware}: if the route $ABCD$ is to be served, the scheduler will never attempt ``downward'' transitions like $A[D]C$. 

The greedy scheduler's advantage lies in the fact that it requires no additional communication infrastructure on top of the one already employed by ebit generation and swapping, since the policy works on strictly local information. The downside to such simplicity is found in the low performance of this policy, that is only interesting as a lower bound for other policies to beat in order to justify the additional communication overhead required. Simulation data for the greedy policy, as well as a comparison with more refined schedulers, is provided in sec. \ref{sec_results}.

\subsection{Quadratic Scheduling}
\label{sec_sub_quadsched}
We now turn to mathematically stating and solving the scheduling problem through the lens provided by our framework. Before solving the problem and displaying results, we introduce our mathematical tools.
\subsubsection{Drift Minimization}
Lyapunov Drift Minimization (LDM) is a standard technique that is often used in classical network science to stabilize queuing systems\cite[sec.\@ 8.4]{MeynTweedieBook}. We provide in this section a demonstration of how and why LDM works, and follow up with its application to quantum networks. 
As a first step, let the Lyapunov Function $V(\mathbf{q}(t),\mathbf{d}(t)) := V(t)$ be an arbitrary, non-negative, convex $\vmathbb{N}^n\xrightarrow{}\vmathbb{R}$ function of the current state of the queues. In short, choosing an arbitrary Lyapunov function and showing it satisfies certain conditions allows to infer that the system is stable. This method entails great simplification of the analysis of highly multivariate systems, because it reduces the problem to a scalar one: when $V(t)$ is small, all the queues are small, and when it is big, at least one queue is accumulating. A common convention\cite{GeorgiadisResAllocation} in network science is to use the square norm of the queue backlog vector as $V(t)$.

After choosing a suitable Lyapunov function, the next step is to define its \textit{drift} $\Delta V(t)$ as: 
\begin{equation}
   \Delta V(t) = \EV{V(t+1) - V(t)|{\mathcal{I}(t)}}.
\end{equation}
We recall that $\mathcal{I}(t)$ is defined as the set of available information at time t, and that we assume all policies presented in this work to have at least knowledge of the state of the queues at time $t$ $(\mathbf{q}(t),\mathbf{d}(t))$.
Some intuition about the drift formulation can be gained by thinking of the Lyapunov function as a potential, akin to the electrical one in physics: the drift is positive if from $t$ to $t+1$ the system evolves into an higher-potential, less stable state, and negative otherwise. 
It is possible to prove\cite[sec.\@ 8.4.2]{MeynTweedieBook} that if $\Delta V(t)$ is negative on the entire state space of the system, except possibly for a compact subset of $(\mathbf{q}(t),\mathbf{d}(t))$ values, then the Markov chain describing the system is positive recurrent, i.e. the network is stable and user requests will not accumulate boundlessly. Such property is known as the Foster-Lyapunov criterion. Intuitively, the drift being positive only on a compact set means that there is a region of the state space in which the system evolves away from stability: since the drift is negative everywhere outside said region the system is always pushed back inside it, so that the Lyapunov function is never allowed to diverge. To visualize this, one may think of a charged particle in a potential well: even if it manages to exit in some way, it is eventually pushed back by the higher potential region.
In its most general form, the Foster-Lyapunov criterion can be phrased as:
\begin{equation}
    \Delta V(t) \leq -f(t) + g(t),
    \label{eq:fosterlyap}
\end{equation}
where $f$ and $g$ are two non-negative functions and the right-hand side is positive on a compact region of the state space of our system. 
Therefore, the practical goal is to find a bound for the drift and minimize it, in order to satisfy the Foster-Lyapunov criterion:
\begin{align}
    \min_{R(t)\in\mathcal{R}}{\Delta V(t)} & \leq -f(t) + g(t) 
    \label{myopicoptimization}
\end{align}
where $\mathcal{R}$ is the set of all feasible scheduling policies.

Notice that everything in our equation is defined only in terms of $t$ and $t+1$: the optimization must be repeated at every time step because of the $t$ dependence, and since the system only sees up to $t+1$ we call this process a \textit{myopic} optimization. Solving the myopic problem at every time step can be proven\cite[appendix]{GiovanidisRACH} to be a suboptimal solution to the infinite horizon Markov Decision Problem of stabilizing the network at steady state.  
\subsubsection{Application to the Framework}
\label{derivation}
We now move to the application of drift minimization to our quantum problem. We first remark that we only seek to stabilize demand queues, because ebit queues play the role of a resource, and their accumulation is not an indicator of the ability of the network to serve user demand (accumulating ebit queues merely amount to more ebits being available and more freedom to the scheduler, especially under unlimited memory assumptions). Additionally, we remark that experimental quantum networks will have a finite number of quantum memory slots at every node, enforcing a hard upper bound on $\mathbf{q}(t)$.

To make our analysis apply to any arbitrary scheduling decision in $\mathbb{N}^n$, we refine our definition of $\mathbf{d}(t)$:
\begin{align}
\mathbf{d}(t+1) &= (\mathbf{d}(t) + \mathbf{b}(t) + \mathbf{\tilde{N}r}(t))^+,
\end{align}
where $(\cdot)^+$ is a shorthand for $\max{(\cdot,0)}$. This is a failsafe measure that prevents the queues in our mathematical model from going negative even if a scheduling policy prescribes more service than there are requests. 

To apply drift minimization to our case, the first step is to choose a Lyapunov function that satisfies the requirements detailed above. As is customary in classical networks, we opt for the square norm of the queue backlog:
\begin{align}
    V(t) = \frac{1}{2}\mathbf{d}^T(t)\mathbf{d}(t).
\end{align}
From there, we obtain the drift:
\begin{equation}
    \Delta V = \frac{1}{2}\EV{\mathbf{d}^T(t+1)\mathbf{d}(t+1) - \mathbf{d}^T(t)\mathbf{d}(t)\big|\mathcal{I}(t)}
    \label{eq:drift_I}
\end{equation}
If we let $\mathbf{d}(t)+\mathbf{b}(t)=\tilde{\mathbf{d}}(t)$ and note that $\left[\max{(x,0)}\right]^2 \leq x^2$ we can bound the drift as: 

\begin{multline}
{
    \frac{1}{2}\EV{\mathbf{d}^T(t+1)\mathbf{d}(t+1) - \mathbf{d}^T(t)\mathbf{d}(t)\big\lvert\mathcal{I}(t)} \leq }\\
    {
    \leq \frac{1}{2}\EV{(\mathbf{\tilde d}(t)+\mathbf{\tilde{N}r}(t))^T(\mathbf{\tilde d}(t)+\mathbf{\tilde{N}r}(t)) - \mathbf{d}^T(t)\mathbf{d}(t)\big\lvert\mathcal{I}(t)} = }\\
    {
    = \frac{1}{2}\left[\EV{\mathbf{\tilde d}^T(t)\tilde{\mathbf{d}}(t)\big\lvert\mathcal{I}(t)} - \mathbf{d}^T(t)\mathbf{d}(t) +
    U(\mathbf{r}(\mathcal{I}(t)),t)\right],
  }
    \label{eq:objderivation}
\end{multline}
where
\begin{align}
{
  U(\mathbf{r}(\mathcal{I}(t)),t):=}&\ {2\EV{\mathbf{\tilde d}(t)\big\lvert\mathcal{I}(t)}^T\mathbf{\tilde{N}r}(\mathcal{I}(t))\ +} \nonumber \\ & \hspace{1cm} {+ \mathbf{r}^T(\mathcal{I}(t))\tilde{\mathbf{N}}^T\mathbf{\tilde N r}(\mathcal{I}(t)).
}
\end{align}
We could pull $\mathbf{d}(t)$ out of the expectation because it is fully determined by $\mathcal{I}(t)$ ($\mathbf{d}(t) \in \mathcal{I}(t)$ for all schedulers by assumption). Furthermore, we have chosen to enforce that the scheduling policies we consider are deterministic and their decisions are completely determined by $\mathcal{I}(t)$, allowing us to also pull $\mathbf{r}(t)\equiv\mathbf{r}(\mathcal{I}(t))$ out of the expectation.

We chose to make the dependence of $U$ on both $t$ and $\mathbf{r}(\mathcal{I}(t))$ explicit to highlight the role of the scheduling decision: whereas stochastic quantities directly depend on $t$, $\mathbf{r}(\mathcal{I}(t))$ behaves as a control parameter: starting from $\mathcal{I}(t)$, the scheduler must tune $\mathbf{r}(\mathcal{I}(t))$ in order to make the controllable part of the drift as negative as possible.
Notice that choosing $\mathbf{r}(\mathcal{I}(t))=0$ leads to $U(\mathbf{r}(\mathcal{I}(t)),t) = 0$: therefore, either the optimal scheduling decision is to take no action, or there is an optimal decision that makes $U$ negative, playing the role of $-f(t)$ in eq. (\ref{eq:fosterlyap}). 

\subsubsection{Fully Informed Quadratic Scheduler}
The derivation presented in the previous section yielded an expression that has a direct effect on stability: the more negative $U(\mathbf{r}(\mathcal{I}(t)),t)$ is, the stabler the network. In other words, the task of a scheduler in this context is to choose at every time step a decision $\mathbf{r}(t)$ such that $U(\mathbf{r}(\mathcal{I}(t)),t)$ is as negative as possible. 

The natural tool to solve this problem is optimization. Assuming, as an initial ideal case, that all information about the network state is available (and therefore dropping the expectation from $U(\mathbf{r}(\mathcal{I}(t)),t)$ since $\mathcal{I}^\text{FI}(t) = \left\{\mathbf{q}(t),\mathbf{d}(t),\mathbf{a}(t),\boldsymbol\ell(t),\mathbf{b}(t)\right\}$, i.e. the realizations of all random variables are exactly known), it is possible to formulate a central scheduling policy that at each time step solves the following quadratic integer program:
\begin{align}
\begin{cases}
\min{\mathbf{w}^{\text{FI}}}(t) \cdot \mathbf{r}(t) + \frac{1}{2}\mathbf{r}(t)^T\mathbf{\tilde N}^T\mathbf{\tilde Nr}(t)\\
\text{s.t. } \mathbf{r}(t)\in{\mathcal{R}^{\text{FI}}(t)}
\end{cases}
\label{quadschedprob}
\end{align}
with weights
\begin{align}
{
\mathbf{w}^{\text{FI}}(t)} = (\mathbf{d}(t) + \mathbf{b}(t))^T\tilde{\mathbf{N}}).
\label{w_fullinfo}
\end{align}

Since we assumed complete information availability, we can use as constraints the feasibility conditions mentioned in sec. \ref{sec:sched_intro} ($d$ being a shorthand for the dimension of $\mathbf{r}(t)$):
\begin{multline} 
    {\mathcal{R}^{\text{FI}}(t)} = \left\{\mathbf{r}(t)\in\mathbb{N}^d
| -\mathbf{\tilde{M}r}(t)\leq 
\mathbf{q}(t) - \boldsymbol{\ell}(t) + \mathbf{a}(t), \right. \\ \left. -\mathbf{\tilde{N}r}(t)\leq 
\mathbf{d}(t) + \mathbf{b}(t)\right\}
\label{quadconstrFULL}
\end{multline}
This constraint set binds the system so that, along every queue:
\begin{itemize}
    \item No more outgoing transitions are scheduled than there are stored ebits;
    \item No more ebits are consumed than there is demand.
\end{itemize}
Solving this problem at every time step will yield a scheduling decision $\mathbf{r}(t)$ that relies on the complete information set, even though such a policy carries a crucial flaw that hinders its experimental realizability: since this is a centralized policy, there must be a physical scheduling block that acts as an authority; all the nodes in the network submit local status information and receive a scheduling decision to apply. In the time it takes for the information to reach the scheduling agent and for the decision to be relayed back to the nodes and applied, the physical layer of the network has continued stochastically generating and losing ebits, so that when the decision finally arrives it is based on outdated information. Two possible solutions to this problem are addressed in the following, in the form of two policies that rely on less information being available. 
\subsubsection{Partially Informed Quadratic Scheduler}
One solution to the stale information problem detailed in the previous section could be to replace all unavailable information with sensible expectation values and thus implement a partially informed quadratic scheduler. We assume that for each queue, the scheduler has access to: \begin{itemize}
    \item The average arrival rate $\alpha$;
    \item The loss parameter $\eta$;
    \item The average demand rate $\beta$;
    \item The system state ($\mathbf{q}(t),\mathbf{d}(t)$) \emph{at the beginning} of each time step,
\end{itemize}
i.e.\@ $\mathcal{I}^\text{PI}(t)=\{\mathbf{q}(t),\mathbf{d}(t),\alpha,\beta,\eta\}$.
This information set relaxes the requirements because the network can take a snapshot of its state at the beginning of each time step and exploit the leftover time to communicate it to the scheduler. The scheduler will in turn use average parameters to build an expectation for the queues' state at the end of the time step and take its decision based on that. Note that if these requirements are still too tight, it is always possible to formulate a policy that knows the exact state of the system with $n$ time steps of delay, or even hybrid localized policies where every node knows the state of the surrounding queues with a delay that depends on their physical distance. 

To formulate our partially informed policy, we re-use the (\ref{quadschedprob}) problem, replacing every quantity for which information is not available with an expected value. Of course, such a heuristic modification degrades the performance of the scheduling policy.
To rely only on information contained in $\mathcal{I}^\text{PI}(t)$, we change the weights of the problem to

\begin{align}
{\mathbf{w}^\text{PI}(t)} = \EV{(\mathbf{d}(t) + \mathbf{b}(t))|{\mathcal{I}^\text{PI}(t)}}^T\mathbf{\tilde N} &\ = \\ = (\mathbf{d}(t) & + \beta\mathbf{1}_{\dim{(\mathbf{d})}})^T\mathbf{\tilde N},
\label{w_partinfo}
\end{align}

where $\mathbf{1}_{\dim{(\mathbf{d})}}$ is the vector of all ones with appropriate dimension, and the constraints to

\begin{align}
\hspace{-.25cm}{\mathcal{R}^\text{PI}(t)} = \left\{\mathbf{r}(t)\in\mathbb{N}^d
\ \middle|\right. & \left.-\tilde{\mathbf{M}}\mathbf{r}(t)\leq 
\EV{\mathbf{q}(t) - \boldsymbol\ell(t) + \mathbf{a}(t)| {\mathcal{I}^\text{PI}(t)}},\ \right. \nonumber\\ 
&\left. -\tilde{\mathbf{N}}\mathbf{r}(t)\leq 
   \EV{\mathbf{d}(t) + \mathbf{b}(t)|{\mathcal{I}^\text{PI}(t)}}\right\},
\end{align}
which in practice reads:
\begin{align}
{\mathcal{R}^\text{PI}(t)} = \left\{\mathbf{r}(t)\in\mathbb{N}^d
\ \middle|\ \right.&\left.-\tilde{\mathbf{M}}\mathbf{r}(t)\leq \eta\mathbf{q}(t) + \alpha\mathbf{1}_{\dim{(\mathbf{d})}}\right., \nonumber\\ 
&\left.-\tilde{\mathbf{N}}\mathbf{r}(t)\leq 
   \mathbf{d}(t) + \beta\mathbf{1}_{\dim{(\mathbf{d})}}\right\}.
\label{quadconstrPART}
\end{align}
This class of partially informed policies still outperforms greedy ones but removes the stale information problem.

It should be stressed that, since this policy relies on a heuristic guess made using averages, it is not guaranteed that its decisions will satisfy the feasibility conditions (conflicts are managed as shown in sec. \ref{sec:conflicts}).

The performance of this policy is reviewed in sec. \ref{sec_results}.
\subsubsection{Node-Localized Quadratic Scheduler}
\label{sec:locquad}
As mentioned before, information availability is one of the main points to consider when choosing a scheduling policy: a well-designed policy must be able to take sensible decisions while leveraging the available information to the best extent possible.

Following this idea, we propose a distributed, optimization-based original policy and subsequently benchmark it to assess its expected performance.

Since we are describing a distributed policy, we shift our point of view to that of a node in the network: we assume that every node $i$ in the network has access to all relevant average values, which can be communicated before the network is booted or measured in a rolling average fashion. 

Additionally, let node $i$ have access to the queue state of the full network at the start of each time step ($\mathbf{q}(t),\mathbf{d}(t)$), where the same remarks we gave in the previous section apply. 

Finally, due to how entanglement generation and swapping are implemented, node $i$ should have access to how many qubits are stored in its memory slots and with whom they are entangled, which means that node $i$ also knows exact arrivals and exact losses for all the queues connected to it, both physical and virtual, and can exploit this additional information when taking a scheduling decision.

To formalize this, let $\mathcal{C}_i$ be the set of queues connected to node $i$, i.e. the set of edges $e$ in the extended set $\tilde{\mathcal{E}}$ such that $e$ is connected to node $i$. Using this concept, we can define a node-local version of the information set $\mathcal{I}^\text{LI}_i(t)$ which contains the entirety of the information available to node $i$:
\begin{equation}
{\mathcal{I}^\text{LI}_i(t)} = \left\{\mathbf{q}(t),\mathbf{d}(t),\eta,\beta,\alpha,a_{e}(t),\ell_{e}(t),b_{e}(t),\ \forall e\in\mathcal{C}_i\right\}\nonumber,
\label{infosetLOC}
\end{equation}

where $a_{e}(t),\ell_{e}(t)$ and $b_{e}(t)$ correspond to the additional local exact information that is unique to each node.

Instead of phrasing a global optimization problem, node $i$ may now formulate an individual problem and solve it to obtain a strictly local scheduling decision to apply directly, without waiting for a discrete scheduler to send back a decision. To do so, the node builds all the relevant quantities (backlogs, arrivals, losses) with exact information from the queues it is connected to and expectation values from the other queues. The $i$-localized quadratic integer program can thus be written as:

\begin{align}
\begin{cases}
\min {\mathbf{w}^{\text{LI}}_{i}(t)} \cdot \mathbf{r}(t) + \frac{1}{2}\mathbf{r}^T(t)\mathbf{\tilde N}^T\mathbf{\tilde N}\mathbf{r}(t) \\
s.t. \ \mathbf{r}(t)\in{\mathcal{R}^\text{LI}_{i}(t)}
\end{cases}
\end{align}
where the weights are given by
\begin{equation}
{\mathbf{w}^{\text{LI}}_{i}(t)} = \EV{\mathbf{d}(t) + \mathbf{b}(t)|{\mathcal{I}^\text{LI}_i(t)}}^T\mathbf{\tilde N},
\label{quadweightsLOC}
\end{equation}

In accordance with its previous definition, the set $\mathcal{R}^\text{LI}_{i}(t)$ of all possible scheduling decisions $\mathbf{r}(t)$ at time slot $t$ localised at node $i$ is defined as:
\begin{align}
{\mathcal{R}^\text{LI}_{i}(t)} = & \left\{\mathbf{r}(t)\in\mathbb{N}^d
\ \middle|\right.\nonumber\\
&\left.-\mathbf{\tilde M}\mathbf{r}(t)\leq 
\EV{\mathbf{q}(t) - \boldsymbol\ell(t) + \mathbf{a}(t)| {\mathcal{I}^\text{LI}_i(t)}},\ \right. \nonumber\\ 
&\left. -\mathbf{\tilde N}\mathbf{r}(t)\leq 
   \EV{\mathbf{d}(t) + \mathbf{b}(t)|{\mathcal{I}^\text{LI}_i(t)}}\right\}.
\label{quadconstrLOC}
\end{align}
In practice, each individual expected value in the weights and constraints' expressions will locally resolve to a form similar to (\ref{quadconstrFULL})  and (\ref{w_fullinfo}) (i.e.\@ all exact values) for queues that are connected to node $i$ and to (\ref{quadconstrPART}) and (\ref{quadweightsLOC}) (all averages) for queues that are not.
As an example, node $A$ will be able to formulate a problem that includes the constraint $-\mathbf{\tilde M}_{AB,} \cdot \mathbf{r}(t) \leq q_{AB}(t) - \ell_{AB}(t) + a_{AB}(t)$ (where $\mathbf{\tilde M}_{AB,}$ is row $AB$ of $\mathbf{\tilde M}$) because queue $AB$ is directly connected to it, but will have to resort to $-\mathbf{\tilde M}_{CD,} \cdot \mathbf{r}(t) \leq \eta q_{CD}(t) + \alpha$ for queue $CD$, because it has no up-to-date information about it.

The locally informed quadratic scheduler provides a practically implementable alternative to the globally informed policy while still retaining good enough performance. We remark once again that, whereas the centralized fully informed method came from abstract mathematical arguments, this scheduler was modified and is thus partially heuristic. Therefore, it is reasonable to expect some degree of performance degradation: one of the tasks of our analysis is to characterize this margin of degradation in order to gauge how close a distributed scheduler can get to its centralized, idealistic variant. 

\subsection{Max Weight Scheduling}
\label{sec_sub_mwsched}
The quadratic policies that have been detailed in the previous section are valid solutions to the scheduling problem in quantum networks. However, situations might arise in which computational complexity is a stricter constraint than network performance. To accommodate such cases, we present in this section a class of policies that perform almost as well as the quadratic ones, for a fraction of the computational cost.

Looking at the policies presented until now, we notice two interesting points: 
\begin{itemize}
    \item The objective function features a linear term that depends on queue length plus a quadratic penalty that does not;
    \item The linear terms are reminiscent of the objective function for the Max Weight\cite{Tassiulas} policy, an extremely well-established result of classical network theory. 
\end{itemize}
It is therefore natural to propose a class of semi-heuristic scheduling policies derived by taking our quadratic objectives and suppressing the quadratic penalty, which does not depend on the queue backlog. For brevity, we explicitly formulate only the fully informed variant of the Max Weight scheduler. The partial and local information quadratic schedulers can be turned to their linear variants following the same steps. 
The fully informed Max Weight problem is obtained by simply suppressing the quadratic term from (\ref{quadschedprob}):
\begin{align}
\begin{cases}
\min {\mathbf{w}^{\text{FI}}(t)} \cdot \mathbf{r}(t)\\
s.t. \ \mathbf{r}(t)\in{\mathcal{R}^{\text{FI}}}(t), 
\end{cases}
\end{align}
and solving it with the same weights and constraints as (\ref{quadconstrFULL}). The partial and local information policies may be similarly constructed by suppressing the quadratic term from the respective quadratic policy. 
The performance analysis for the globally, partially and locally informed linear schedulers is provided in section \ref{sec_results}.

\section{Numerical Analysis}
\label{sec_numerical}
In this section, we give an overview of how our simulation tool works and then provide results for the numerical analysis of all the proposed schedulers.
\subsection{Simulator Architecture}
All the results shown in this work were obtained through an ad-hoc simulator implemented in Python, relying on the \texttt{gurobi}\cite{gurobi} solver for the optimization calculations and \texttt{networkx}\cite{networkx} as a graph backend. In the following, we provide a quick breakdown of how our simulator works, from the point of view of a user that is not necessarily experienced with writing code. Interested readers may find more information on the simulator's GitHub repository \cite{gitrepo}. 

From a black-box perspective, the focus of the code design phase of our work was on an object-oriented model of the network system that is as modular and layered as possible. The motivation driving this approach was that an ideal version of the controlling code should be abstract enough not to be aware whether it is driving our model, another more refined simulator or even a real network. In the following, we give a brief rundown of the kind of parameters that a user of our framework and simulator may expect to tune. 

The simulator's input files are composed of two sets of ingredients for the user to provide: the first set of parameters is devoted to the generation of the network topology, the choice of service pairs and demand rates. Users are free to choose one of the topologies we propose in this work (with tunable parameters) or provide an entirely custom network graph. 

After selecting the topology, the user selects the set of scheduling policies that the simulator will analyze. As before, it is possible to select one of the policies we analyzed here or provide a custom one. The code provides seamless access to all the information we used in our policies through simple specification of an ``information availability'' parameter. 
 
The second set of input values is related to physics and low-level simulation parameters, enabling fine-tuning of generation rates across physical links and losses at nodes, but also number and duration of the time steps. 

A set of parameters related to the optimization of the simulator's performance concludes the user inputs for our code. A discussion of these parameters is out of the scope of this paper as they are only relevant to raw computational performance, but can be found in the full code documentation of the simulator on GitHub\cite{gitrepo}.
\subsection{Results}
\label{sec_results}
\begin{figure*}[b]
\centering
\subfloat[Fully informed Max-Weight;]{\includegraphics[width=.28\textwidth]{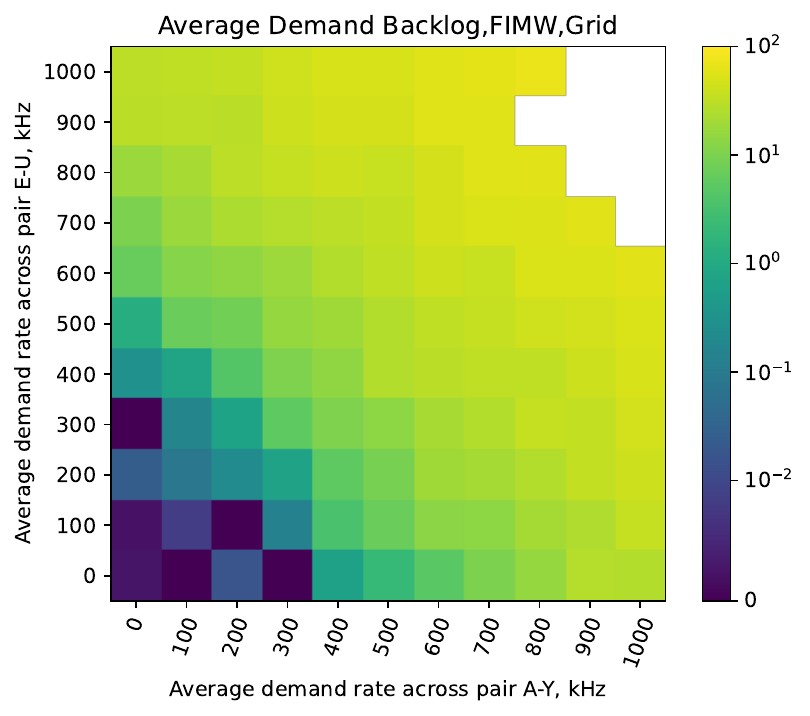}}
\subfloat[Locally informed Max-Weight;]{\includegraphics[width=.28\textwidth]{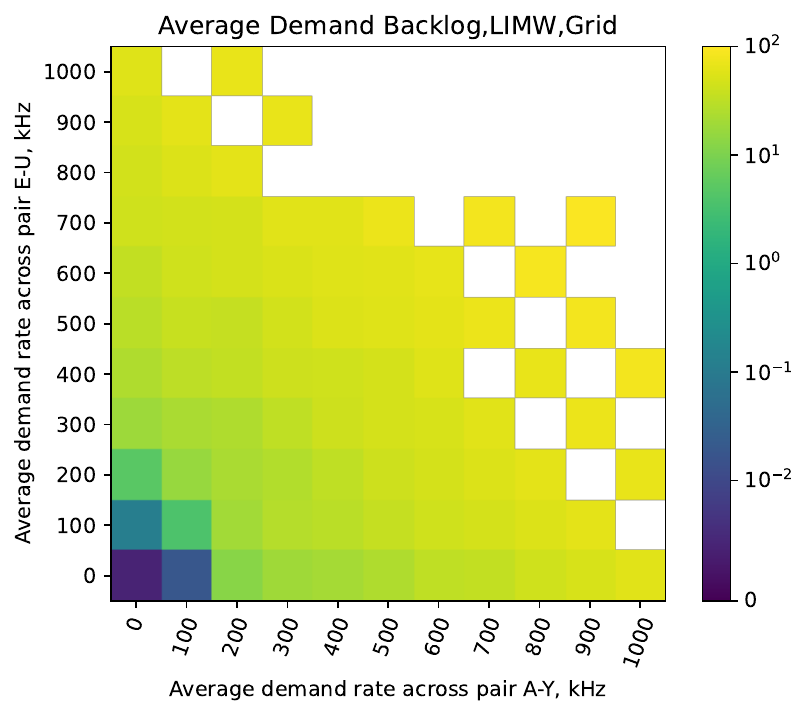}}
\subfloat[Average-only Max-Weight;]{\includegraphics[width=.28\textwidth]{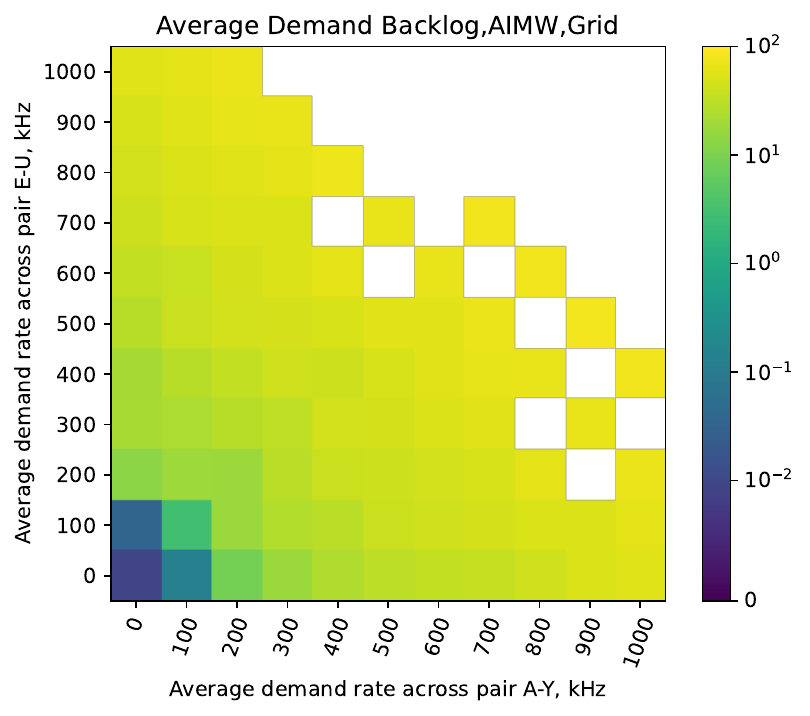}}

\subfloat[Fully informed Quadratic;]{\includegraphics[width=.28\textwidth]{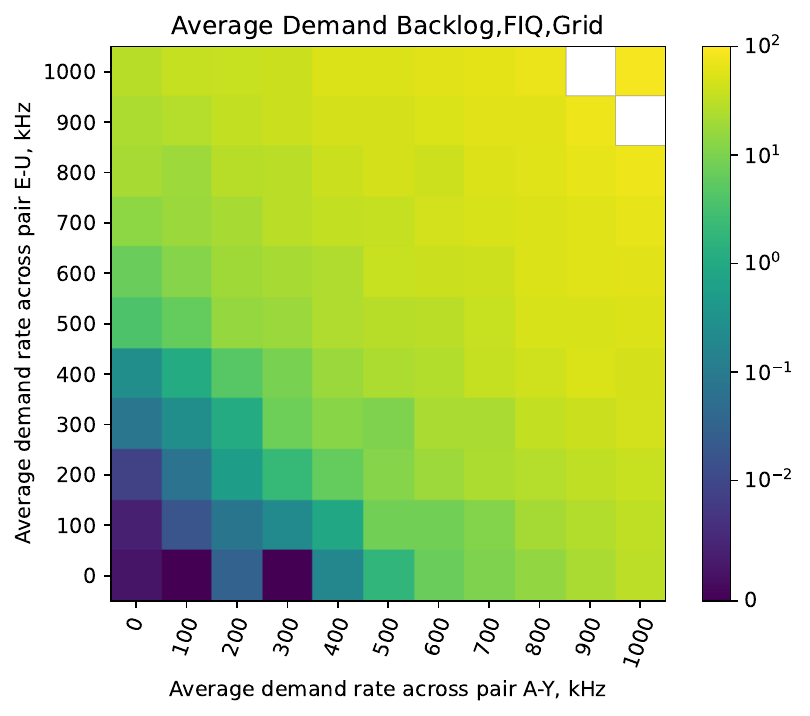}}
\subfloat[Locally informed Quadratic;]{\includegraphics[width=.28\textwidth]{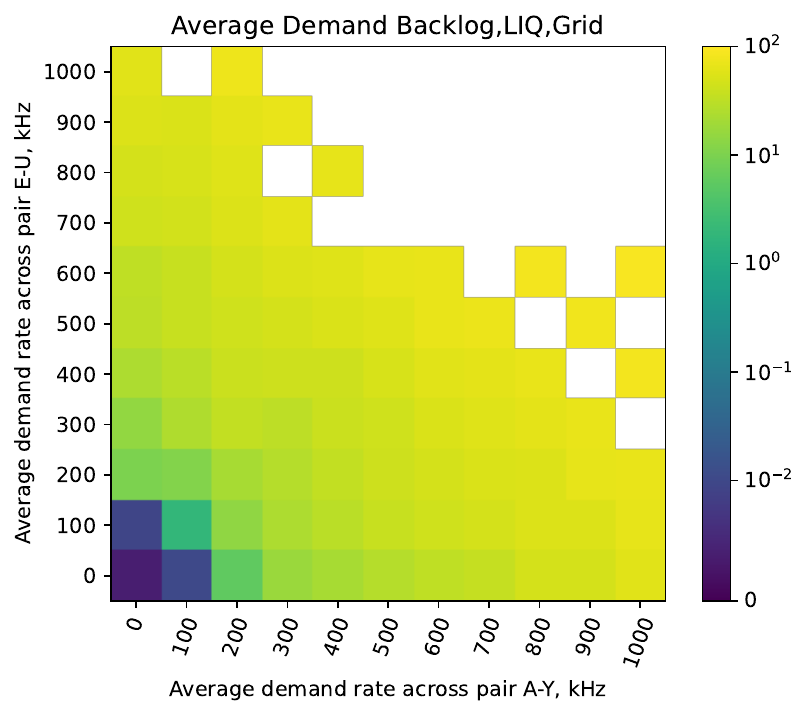}}
\subfloat[Average-only Quadratic.]{\includegraphics[width=.28\textwidth]{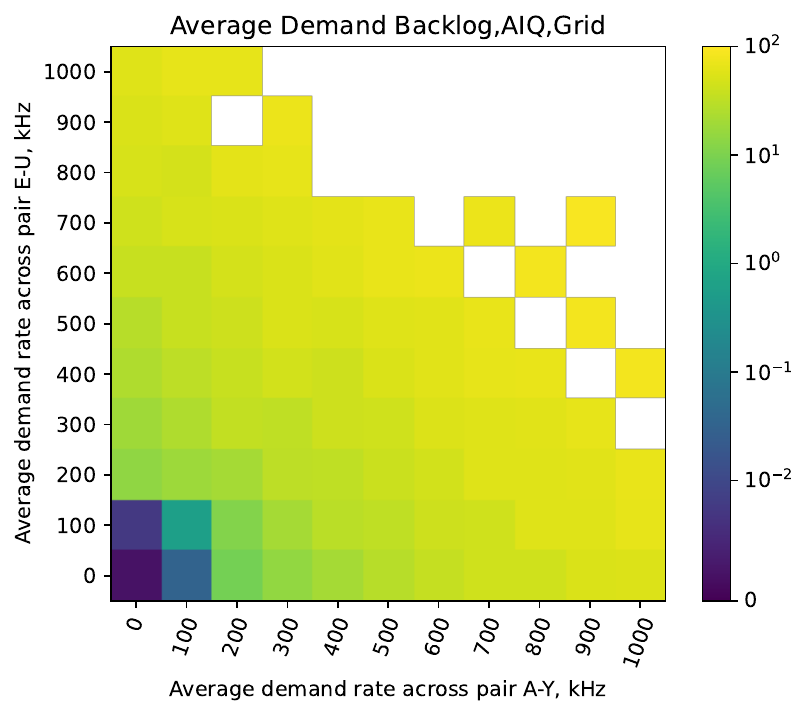}}
\caption{Comparison of the performance of the linear scheduling policies we presented in the main text and their quadratic counterparts. For brevity, we only report results for the Grid topology shown in fig. \ref{fig:gridtopo}, while stating that the same phenomenon is observed for all topologies: the margin of performance between Max-Weight and Quadratic schedulers is almost imperceptible in our tests. This figure was calculated with an additional set of eight random parasitic pairs, whose average load was fixed at $100$ kHz. Analogously to the main results in fig. \ref{fig:stabgrids}, simulations were run for $1000$ time steps of $1 \mu s$, discarding the first $100$ observations to reduce the impact of transients. 
The white points were skipped by the simulator and directly deemed unstable, since one or more strictly lower-load points were found to be unstable. More details on this computational economy technique can be found in the main text.}
\label{fig:quadperf}
\end{figure*}

To avoid excessively prolonging this section, we show in fig. \ref{fig:quadperf} that the quadratic schedulers provide a negligible, if any, increase in performance at the cost of a major increase in computational complexity (quadratic optimization calculations are much more taxing than linear ones). They were therefore omitted from the complete discussion of numerical results, that only shows the greedy scheduler and the three linear ones. 

\begin{figure*}
\subfloat[A complete 5x5 grid;]{
\includegraphics[width=.5\textwidth, trim =53 0 53 20, clip]{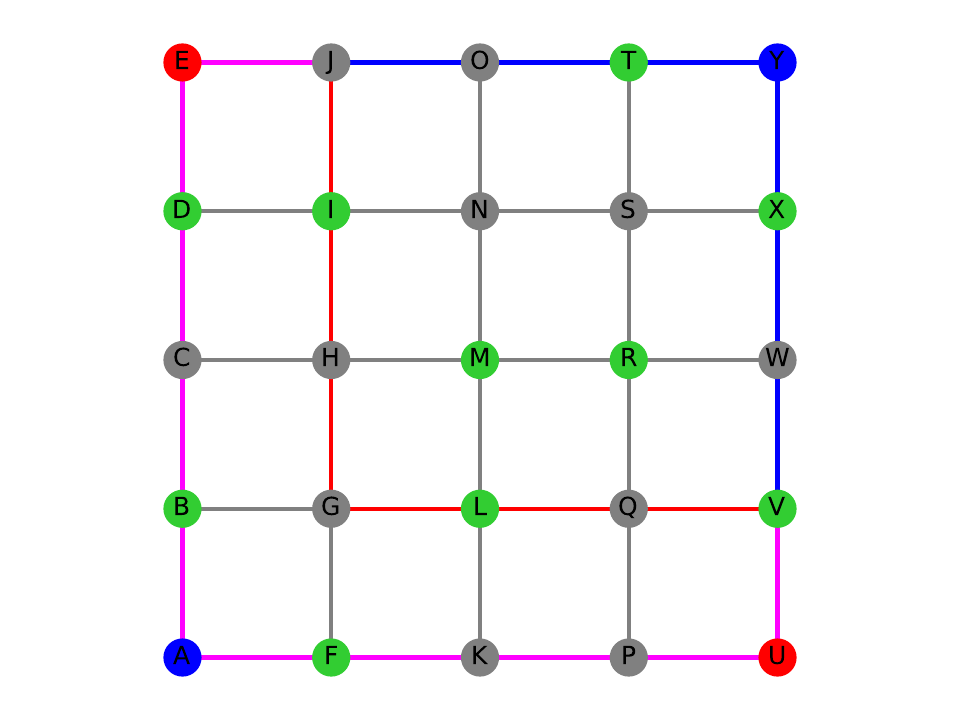}
\label{fig:gridtopo}
}
\hfill
\subfloat[A 6x6 grid whose nodes had a probability $p=0.25$ of being removed;]{\includegraphics[width=.5\textwidth]{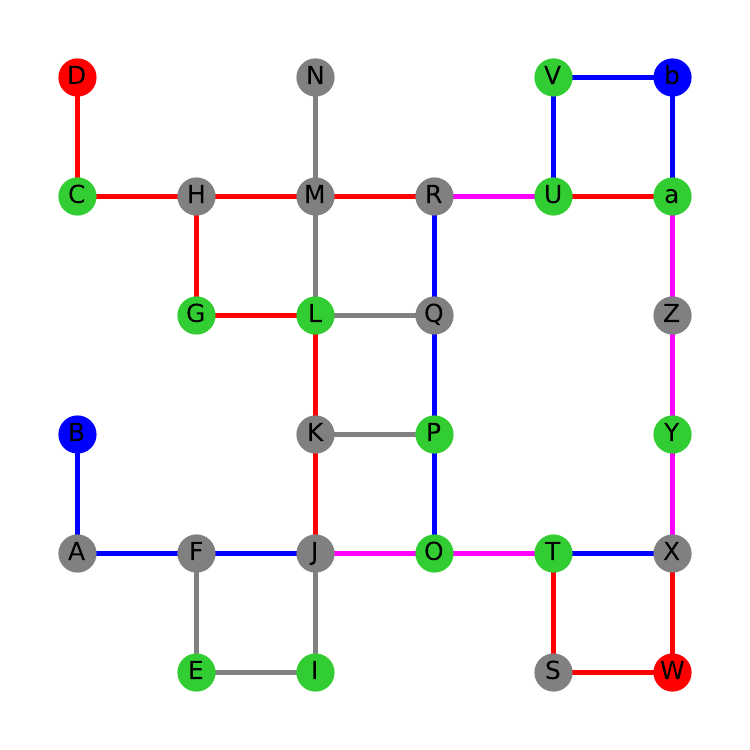}\label{fig:pgridtopo}}
\hfill
\subfloat[An Erdős–Rényi random graph, with $n=25$ and $p=0.125$;]{\includegraphics[width=.5\textwidth, trim =145 145 145 145, clip]{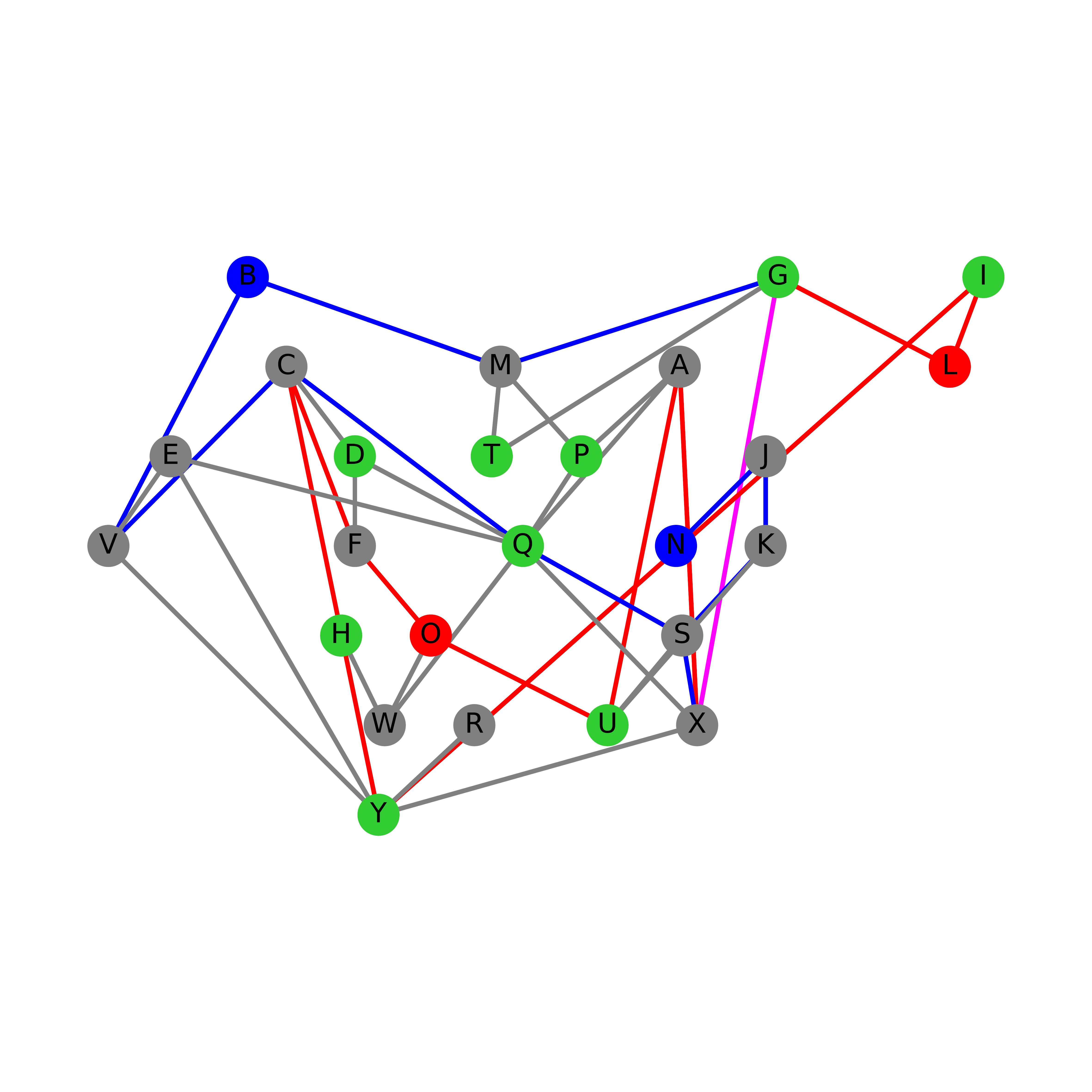}\label{fig:ERtopo}}
\hfill
\subfloat[A Watts--Strogatz random graph, with $n=25$, $n_{\text{neighbors}} = 4$ and $p=0.2$.]{\includegraphics[width=.5\textwidth]{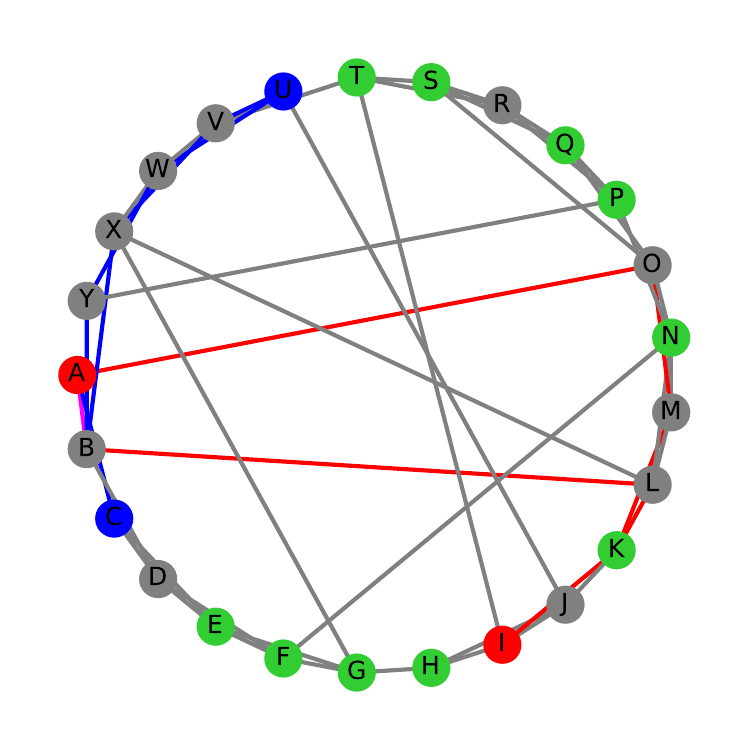}\label{fig:WSTopo}}
\caption{The four topologies analyzed in this work. The main service pairs and the routes connecting them have been highlighted in red and blue, with purple representing shared edges, i.e. edges that appear in both pairs' service routes. In green, we provide a visual example of the random parasitic pairs: every green node is paired with another colored node and requests entanglement with a fixed rate. At every run of the simulator, we redraw the green pairs to study the effect of traffic without bias towards a specific configuration.}
\label{fig:topos}
\end{figure*}

The main goal of the following analysis is to showcase how the proposed scheduling policies affect the performance of quantum networks of various topologies, both deterministic and randomly generated. The topologies on which our analysis was run, shown in fig. \ref{fig:topos}, are a complete 5x5 grid, a 6x6 grid with some randomly removed nodes, and two realizations of the Watts--Strogatz\cite{WattsRandom} and Erdős–Rényi\cite{ErdosRandom} models of 25 nodes each.

Since our $\mathbf{M}$ matrix is built from the static routes that connect the service pairs, building a nontrivial example requires more than two routes. To obtain such an example, we increase the number of users we consider: for each topology, we run our simulation with ten user pairs, of which two are manually fixed (red and blue in fig. \ref{fig:topos}) and eight are randomly selected at the beginning of each simulation run to mimic different traffic configurations (green in fig. \ref{fig:topos}). Every user pair is connected, when possible, by two semi-distinct routes. Since routing is outside the scope of this work, for we simply take the shortest path connecting each user pair, remove the edges that compose it with a given tunable probability, and then take the shortest path in the newly obtained graph as a second route, under the assumption that in a real application scenario users will provide sensibly computed static routes. 

We sweep the demand rate of the two manually selected pairs, while fixing the random ones to a constant load value $L$, and then average together the results of ten runs to remove the bias that one particular parasitic pairs set may entail. For all individual runs exploited to calculate the results we present in this section, the simulator was run for $1000$ time steps of $1 \mu s$ each, discarding the first $100$ steps to reduce the impact of transients.

\begin{figure*}[ht]
\centering
\includegraphics[width = .9\textwidth]{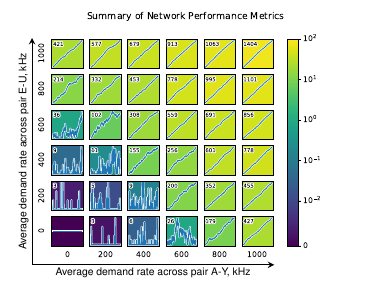}
\caption[Global summary of the network performance metrics.]{Global summary of all the network performance metrics that can be analyzed through our simulator, when running the full information Max Weight scheduler over the grid topology from fig. \ref{fig:gridtopo}. 

Inside each cell:
\begin{enumerate*}
    \item The plot shows the temporal evolution of total demand from start to finish; 
      it allows to easily distinguish stable regime (with finite excursion) from the unstable one 
      (with a linear trend);
    \item The background color represents the average demand backlog throughout a simulation run; 
    \item The top-left number is the maximum excursion of the total demand in the network;
      in the stable regime, it can be seen as a rough upper bound on the amount of quantum memory
      required at each node to achieve this performance level.
\end{enumerate*}
}
\label{fig:TimePlots}
\end{figure*}

Fig. \ref{fig:TimePlots} provides a showcase of all the results that we obtain from our simulation: given the complete 5x5 grid topology shown in fig. \ref{fig:gridtopo} and the fully informed Max Weight scheduler, we select the four corners of the grid as the two main user pairs, randomize the parasitic pairs and run the simulation, displaying all outputs.

Since tracing the capacity region of a network requires gauging its stability, we rely on fig. \ref{fig:TimePlots} as an aid to clarify our definition of this crucial concept.
In the context of dynamical systems, stability may be defined in several different ways, depending on the amount of mathematical rigor required. The established definition states that a system of queues is stable if the  time it takes for the cumulative queue length to return to zero is finite on average (i.e.\@ the queues keep returning to zero, regardless of how far they stray). Of course, such a definition loses meaning in a finite-time context, because there is no way to tell whether a system would turn back to zero if left running for a longer wall time, even though it looks unstable over a finite time window. 
However, arguments can be made to justify the usage of such a notion in a context such as ours. First of all, it is safe to say that a queue whose length is constantly zero is stable (This is apparent from fig. \ref{fig:TimePlots}, plot in the $(0,0)$ cell, which depicts the temporal trend of the total demand, with all demand rates set to zero). Secondly, we may state that a queue that has Poissonian arrivals and is never depleted will accumulate in a roughly linear fashion, and it will surely be unstable. Thirdly, we claim that the stability front of a network system is a Pareto boundary: if a given load $L = (l_1,l_2,...l_i,...,l_n)$ cannot be served by the network and is therefore outside its stability region, then all higher loads $L' = (l_1,l_2,...l'_i,...,l_n)$ such that $l'_i > l_i$ are unstable (fig. \ref{fig:TimePlots}, upper-right cluster of linear plots, depicting total demand in a high-load scenario). 

These considerations make a finite-time simulation slightly more insightful: if the queue length returns to zero several times during the simulation window, the system is likely stable. If the system shows a clear linear trend, there is high possibility that it is not. If a cluster of points all show a linear trend, the possibility of instability further increases. 

Moreover, to conform with standard practice in the classical network field, we also include as a performance metric the average demand queue length, plotted as a colormap in the background of fig. \ref{fig:TimePlots}'s cells. This is the metric on which we focus for the rest of the analysis, since it yields a more easily legible graph of the stability of a load point and is therefore more suitable for high-resolution plots and/or comparison of a large number of results. Another reason why we choose to present the average queue length as a color map is that it provides a visual approximation of the capacity region of the network we are considering.

To give a sense of scale, we complement our outputs with the maximum excursion of the cumulative demand backlog, shown in the top-left corner of every cell.

\begin{figure*}
\subfloat[Complete grid;]{
\includegraphics[width = .5\textwidth]{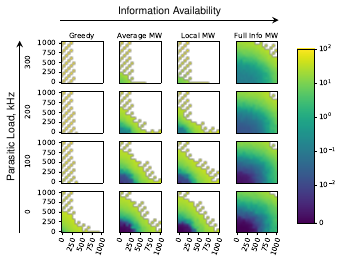}
\label{fig:finalgrid}}
\hfill
\subfloat[Grid with probabilistically removed nodes;]{
\includegraphics[width = .5\textwidth]{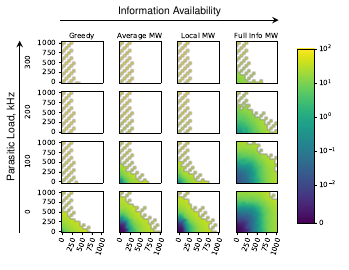}
\label{fig:finalholedgrid}}

\subfloat[Erdős–Rényi random graph]{
\includegraphics[width = .5\textwidth]{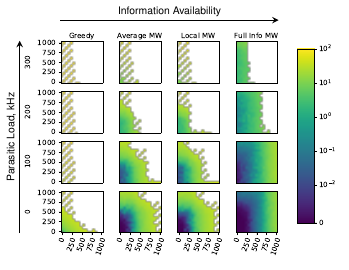}
\label{fig:finalER}}
\hfill
\subfloat[Watts--Strogatz random graph.]{
\includegraphics[width = .5\textwidth]{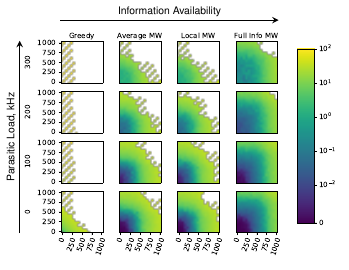}
\label{fig:finalWS}}
\caption{For each of the four topologies, we provide a grid of plots obtained by simulating different operating points. As mentioned in the main text, there are ten pairs of users, of which two are fixed and eight randomized. Each cell of the grids is a plot of the average demand backlog vs.\@ the load across the two main pairs (reported in $kHz$ on the small axes of the individual cells) under certain operating conditions. The conditions in which every plot was calculated are fixed by the Information Availability and Parasitic Load meta-axes, the former indicating which scheduler was employed to control the network (Greedy to Full Information, in increasing order of available information), the latter the load placed upon the randomized parasitic pairs in $kHz$. As discussed in the main text, a dark blue point is deemed stable and a yellow one unstable, while the middle grounds are somewhat ambiguous due to the finite-time nature of the simulation. 
The white points have not been calculated by the simulator to save time, since a point at a lower load was found to be unequivocally unstable and the stability region is expected to be a Pareto bound. We recall that every cell in the grids comes from averaging ten different traffic configurations, where a configuration consists of the same two main pairs and a fresh set of eight parasitic ones. The shape of each stability region may be seen as a measure of competition between user pairs: the more diagonal the boundary of the dark blue region, the higher the direct competition between the main pairs. The difference in area of regions along one given column is a direct measurement of how the main and parasitic pairs compete (and therefore how the network serves requests under increasing stress), while the differences along one row show how well the scheduler leverages additional information.}
\label{fig:stabgrids}
\end{figure*}

Running our analysis over all topologies and schedulers and displaying the average demand backlog, we obtain four arrays of plots that show the performance of our network as a function of the information granted to the scheduling policy (Greedy to Fully Informed Global) and the load on the parasitic pairs, shown in fig. \ref{fig:stabgrids}. From these arrays of plots, insight on several levels may be obtained. 

Firstly, looking at all the plots for any given topology, we observe that changing the scheduler entails noticeable change on the capacity region of a quantum network, providing proof that not only the scheduling problem is an interesting one to formulate in the context of quantum networking, but its solution brings non-negligible performance margins to the operation of a quantum network. Another piece of information that may be gathered resides in the shapes of the stability margin: when the deep blue region is not shaped like a rectangle it means that the two plotted pairs are in direct competition, as increasing demand along one of the axes reduces the amount of demand that can be served along the other one. To an end user employing our tool for network design, this would mean that the network is bottlenecked by routing, since there is a set of routes across which the scheduler must balance service to two or more competing commodities.

Another point that can be made from the results in fig. \ref{fig:stabgrids} comes from looking at the difference between the fully informed global scheduler and the local ones: as mentioned before, the fully informed Max Weight scheduler can be interpreted as a performance upper bound for a Max Weight policy. Therefore, when designing an original scheduler, one may gauge its performance by comparing stability regions with the fully informed scheduler. There is a noticeable difference between FI and LI but it may be deemed acceptable because of the information trade-off: the region still has the same shape and, although smaller, is still comparable to the upper bound, meaning that the locally informed policy we are proposing performs very well in this scenario. 

When the stable section of the service region features a diagonal bound (fig. \ref{fig:finalgrid}, Fully Informed scheduler, $0$ load), increasing demand across one of the main pairs directly impairs service across the other, signaling a bottleneck over which the two pairs are competing.  Conversely, a rectangular shape (e.g.\@ fig. \ref{fig:finalER}, Fully Informed scheduler column) is an indicator that the two main pairs we selected are not directly competing over a shared bottleneck. This does not necessarily mean that the network is not congested: traffic from the parasitic pairs is still stressing the network (as demonstrated by the reduction in size of the stability region when going up along the parasitic load axis) and requiring careful scheduling decisions.

\section{Limitations of the Framework and Future Outlook}
In this section, we discuss the main limitations and open questions in our model, and propose some seed ideas for future directions. The first limitation to talk about is the modelization of strictly quantum imperfections such as decoherence, that degrade the quality of a quantum state without necessarily meaning the state is lost. Despite being well aware of the paramount importance of noise in quantum modeling, the history of the classical Internet shows that a successful large-scale network infrastructure is best thought of in terms of separate functional layers, and a layered architecture has already been proposed for a prospective future quantum internet \cite{QStack} that effectively separates the Link Layer, where quantum error correction should be implemented, from the Network Layer, which is the scope of our work. While we are aware that in real implementations, especially initial ones, theoretically separate layers leak and blend with each other, the Quantum Internet should eventually converge to a well defined network stack, making it redundant to treat noise in the same layer as scheduling. Thus, while we remain interested in an expansion of our work that treats quantum imperfections, the lack of explicit state quality modeling does not make our work irrelevant.

A similar concern could be raised for the memory at the network nodes: despite this being another issue that is very close to hardware, its integration with scheduling policies would seem crucial because it could intrinsically change how a scheduling decision is taken: if a node only has a finite number of memory slots, the scheduler would have the additional constraint of free space (or lack thereof, in some cases having to ``waste'' ebits in order to free up memory). As a matter of fact, a similar problem has been analysed over a single switch in \cite{VardoyanSwitchStochastic} and \cite{NainSwitch}, showing that the memory requirements of an isolated quantum switch are quite low (on the order of 5 slots) to achieve performance comparable to that of a switch with unlimited memory slots, making the memory problem not as concerning. Moreover, \cite{PromponasMemory} formulates the problem of exploiting limited memory slots and develops a Max-Weight memory allocation policy for quantum nodes that could be adapted to our scenario. 

Furthermore, it is possible to look at the memory problem from a different direction: while a solution inside our framework could in principle be to add compound constraints to the optimization problems, we stress that results such as fig. \ref{fig:TimePlots} (maximal excursion numbers) gauge the accumulation of total demand in a stable network, effectively providing an upper bound for memory requirements in the design of a real quantum network system.

The third limitation of our work is how the framework scales: The fact that the number of queues we need to account for grows quadratically with the number of nodes in the network entails quick growth of the $\mathbf{M}$ matrix, which makes the integer programs required by several policies presented here increasingly complex. While this is not as much of a problem currently as it was in the past decades, it is still an issue that is worth closely investigating, perhaps to find scheduling strategies that require only a subset of the extended edge set (akin to an overlay network, as demonstrated in \cite{Pouryousef2022QON}). We note here that easing scaling concerns would also enable a future extension of our framework to multipartite entanglement: as mentioned in the beginning, an extension in this direction would require the definition of new multipartite virtual queues, together with ad-hoc transitions that interface them with the bipartite ones, greatly increasing the overall number of queues and therefore the problem's complexity.

Moreover, this work does not provide analytical proofs of Lyapunov stability or optimality of the proposed families of policies, which are of great interest in network science and could be promising directions for future work. To provide a starting point, we direct the interested reader to the well-known optimality results of Max Weight on classical networks \cite{TassiulasMaxWeight} and to \cite{VasantamSwitch}, where Lyapunov stability and throughput optimality are analytically proven for a Max Weight policy in the case of a switch without quantum memory serving three users in a star topology: it is possible to show, by translating the referenced paper's model into our framework, that the Max Weight policy presented in the cited work is equivalent to the Fully Informed Max Weight analyzed in ours. 
Furthermore, we stress that in the case proposed by \cite{VasantamSwitch} our class of Quadratic policies reduces to Max Weight: since \cite{VasantamSwitch} employs Bernoullian ebit arrivals and no quantum memory, the components of the scheduling decision $\mathbf{r}(t)$ can be at most $1$. Coupled with the structure of $\mathbf{\tilde{N}}$, this entails that at all time steps the quadratic penalty has the same value across all possible scheduling decisions, reducing the more general Quadratic objective function to a Max Weight one and showing that in this special case our policies are optimal, which offers potential for investigation as to how generalizable the optimality claim is. We note that, since we included additional information in the Lyapunov drift definition, a formal proof of stability would also require averaging over all the additional environmental factors that were included in the conditioning of \ref{eq:drift_I} other than the queue state ($\mathbf{q}(t),\mathbf{d}(t)$).

Finally, it would be interesting to delve into other physical imperfections, such as finite speed of communication between nodes, which entail a stricter definition of what information is local and accessible to a node at a given time. One interesting implication of such analysis would be the case in which only one of the qubits in an ebit is lost, and what happens if the loss is not communicated before other swapping operations are undertaken, i.e.\@ the error propagates along the swapping route. 
All these considerations would require a more refined physical model, which would in turn imply revisions to our mathematical framework, but should not be excessively difficult to include in the numerical part of our discussion: the simulator code was written from the ground up in order to provide a simpler and more agile contribution, but it was designed with particular attention to keeping a layered and modular structure that should be reasonably adaptable to well-established quantum network simulation packages such as NetSquid\cite{CoopmansNetSquid} or QuISP\cite{SatohQuISP}.

\section{Conclusions}
In this work, we presented a general framework that allows to formulate and solve the scheduling problem in general, lossy memory-endowed quantum networks in a dynamical way. We then integrated our framework with Lyapunov Drift Minimization in order to mathematically derive a throughput-efficient quadratic scheduling policy for quantum networks and proposed several other heuristic policies with various advantages. Finally, we showcased how our framework may be exploited by people interested in policy design to benchmark and fine-tune a general quantum network's performance under arbitrary scheduling policies. Despite a sizable amount of work still needing to be tackled before a collective quantum network science exists, the promising results we presented could eventually become one of many assets in the quest for the Quantum Internet.
\bibliographystyle{IEEEtran}
\bibliography{IEEEabrv,bibliography}
\EOD
\end{document}